\documentclass[a4paper,amsmath, 12pt]{article}

\usepackage{color}
\usepackage{graphicx}
\usepackage[a4paper]{geometry}
\usepackage{amsmath}
\usepackage{rotating}
\usepackage{lscape}
\usepackage{subcaption}
\usepackage[symbol]{footmisc}
\usepackage{multirow}
\usepackage{threeparttable}
\usepackage{natbib}
\usepackage{xcolor}
\usepackage{hyperref}
\usepackage{subcaption}

\begin{document}

\begin{center}
\textbf{\Large Bivariate copula regression models for semi-competing risks }

\vspace{0.5cm}

Yinghui Wei$^{1}$\footnote[1]{Corresponding author: Yinghui Wei, Centre for Mathematical Sciences, School of Engineering, Computing and Mathematics, University of Plymouth, PL4 8AA, UK. Email: yinghui.wei@plymouth.ac.uk.}, Ma{\l}gorzata Wojty{\'s}$^1$, Lexy Sorrell$^1$
and Peter Rowe$^2$

{\it $^1$Centre for Mathematical Sciences, School of Engineering, Computing and Mathematics, University of Plymouth, PL4 8AA\\

\it $^2$South West Transplant Centre, University Hospitals Plymouth NHS Trust, PL6 8DH}
\end{center}

\begin{abstract}
\noindent Time-to-event semi-competing risk endpoints may be correlated when both events are occurring on the same individual. These events and the association between them may also be influenced by individual characteristics. In this paper, we propose copula survival models to estimate hazard ratios of covariates on the non-terminal and terminal events, along with the effects of covariates on the association between the two events. We use the Normal, Clayton, Frank and Gumbel copulas to provide a variety of association structures between the non-terminal and terminal events. We apply the proposed methods to model semi-competing risks of graft failure and death for kidney transplant patients. We find that  copula survival models perform better than the Cox proportional hazards model when estimating the non-terminal event hazard ratio of covariates. We also find that the inclusion of covariates in the association parameter of the copula models improves the estimation of the hazard ratios.

\end{abstract}

Keywords: copula model, renal transplant, semi-competing risk, survival analysis, hazard ratio

\section{Introduction}

Often in medical studies, patients who are lost to follow-up, or do not experience the event of interest during the study period, leave a censored observation. However, with semi-competing risk endpoints, the non-terminal event has the possibility of also being censored by the terminal event \citep{Fine2001}. One example of semi-competing risks, which will be analysed later in this paper, is graft failure and death following renal transplant, where the non-terminal event is graft failure and the terminal event is death.

\noindent The Cox proportional hazards model is widely used in practice to analyse time-to-event outcomes. The Cox model was originally developed to analyse all-cause mortality \citep{Cox1972}, of which there is no competing risk. The hazard ratios estimated from the Cox model, assuming an independent censoring mechanism, are potentially biased when analysing a non-terminal event \citep{Berry2010}. Some authors use copula regression models to jointly model non-terminal and terminal events. \cite{Peng2007}, \cite{Hsieh2008} and \cite{Chen2012} propose semiparametric copula models for the regression on the marginal distributions. The analysis on the terminal event can be conducted using common survival methodology. For instance, \cite{Peng2007} model the terminal event with a proportional hazards model marginally within their copula model. \cite{Hsieh2008} use a copula model, which allows for different dependence structures between covariate groups. The Bayesian Normal induced copula estimation model is developed by \cite{Fu2013}. \\

\noindent  Copulas are multivariate distribution functions which can model the marginal distributions separately, along with the dependence structure between them. \cite{Sorrell2022} introduced bivariate copula models to estimate the correlation between semi-competing risk endpoints, using the following four copula functions, the Normal, Clayton, Frank and Gumbel copulas. However, the hazard rates of the non-terminal and terminal events, along with the association parameter between them, may be influenced by covariates. Including covariates into the analysis of the correlation between survival endpoints can help understand how the association may be influenced by individual characteristics. Covariates may also be included in the analysis of marginal distributions, which allows to estimate  hazard ratios and subsequently the comparison of risks of survival endpoints between groups.  \\

\noindent In this paper, we develop copula survival regression models by using conditional copula \citep{Patton2006b}, which allow both the association parameter between the survival endpoints and the hazard rates to depend on multiple binary covariates. This is also an extension from the copula survival model introduced in \cite{Sorrell2022}, where no covariates are included. We estimate the hazard ratio using copula survival regression models with binary covariates. We also estimate the effects of these binary covariates on the association between semi-competing risks. By jointly modelling the non-terminal and terminal events and including the correlation between them, we hope to improve the inference about the non-terminal event. \\

\noindent We focus on the inclusion of covariates in the analysis of survival data with a semi-competing risk using conditional copulas. In the semi-competing risk setting, \cite{Chen2012} developed a model allowing the association parameter of the copula function to vary with categorical covariates. A model allowing each covariate group to assume a separate Archimedean copula function is introduced by \cite{Hsieh2008}. This allow to have different dependence structures describing the association between the semi-competing risk events for each covariate group. The effect of a discrete covariate on the non-terminal and terminal event and the association parameter between the semi-competing risk events is investigated by \cite{Ghosh2006}. \cite{Peng2007} studied the effect of a covariate on the non-terminal event and the association parameter, by proposing a model with a time-dependent copula and apply the methods to survival endpoints following AIDS diagnosis. Similarly, \cite{Zhu2021} propose a copula regression approach to examine covariate effects on the non-terminal and terminal events using a two-stage approach. In stage 1, the covariate effects on the marginal events are investigated and in stage 2, the association parameter of the copula model is estimated. \cite{Zhu2021} prefer the two-stage approach, compared to the one-stage approach by \cite{Peng2007}, as it allows for separate identification of misspecification of the marginal regression models and the copula model.

The rest of the paper is organised as follows. The motivating data for this paper are described in Section \ref{sec:data}. We then introduce our proposed Copula regression survival methods to estimate the effect of covariates on the semi-competing risk events and on the association parameter using a variety of conditional copula functions in Section \ref{sec:meth}. In Section \ref{sec:analysis}, we apply our proposed methods to the motivating data, followed by a simulation study to compare the copula models in Section \ref{sec:sim}. We conclude with a discussion.

\section{Motivating data} \label{sec:data}

The motivating data are from the United Kingdom Transplant Registry (UKTR), held by the National Health Service (NHS) Blood and Transplant. The outcomes, time to graft failure and time to death since kidney transplantation, are semi-competing risks. Our study population is kidney transplant recipients, who had their single and first transplant between 1995 and 2016 in the UK, with known age, sex and donor type at the time of transplantation. We present a novel application by using our proposed copula survival regression models.

We included $40,348$  single and first kidney transplants between 1995 and 2016 in the UK. For $78.01\%$ of patients the graft failure time is  censored, for $80.0\%$ of patients the death time is censored and for $64.3\%$ of patients both events are censored.

Table \ref{table:NHSBT_baseline} shows the baseline characteristics of interest. The covariates to be included in the application later in this paper include donor type (deceased or living), recipient's sex (male or female), and recipient's age ($>50$ years or $\le 50$ years). Donor type indicates whether a recipient had received a deceased donor or living donor kidney for their transplantation.
These covariates were included as they are important factors influencing transplant outcomes according to the literature as follows. As reported in \cite{Terasaki1995} and \cite{Port2004}, living kidney donor recipients have improved survival compared to deceased donor recipients.

Moreover, recipient's age has been shown to be a significant factor affecting survival following transplant \citep{Becker2000,Fabrizii2004,Keith2004}. In \cite{Fabrizii2004}, the 5-year survival was found to be affected by recipient age group, however, these differences were not evident after controlling for confounding.

\begin{table}[h]
\centering
\begin{tabular}{|l l | l|}
\hline
\multicolumn{2}{|l|}{\textbf{Variable}} &${n \, (\%)}$ \\
\hline
\multirow{2}{*}{\textbf{Donor type}} & Deceased & $27,971$ ($69.9\%$) \\
                   & Living        & $12,377$ ($30.1\%$) \\
\hline
\multirow{2}{*}{\textbf{Recipient's sex}}
                      & Male     & $24,945$ ($61.8\%$) \\
                      & Female & $15,403$ ($38.2\%$) \\

\hline
\multirow{2}{*}{\textbf{Recipient's age (years)}}
                    & $\le 50$         &  $23,208$ ($57.5\%$) \\
                    & $>50$       & $17,140$ ($42.5\%$) \\

\hline

\end{tabular}
\caption{Baseline characteristics for single and first kidney transplant recipients from the UK Transplant Registry data set in (1995 $-$ 2016). The number of recipients with the particular characteristic is given alongside the percentage in brackets.}
\label{table:NHSBT_baseline}
\end{table}

\normalsize

\section{Methods}\label{sec:meth}

In this section, we describe methods to include covariates in copula functions to examine the effects that they have on the semi-competing risk events. We let the association parameter of the copula function, along with the hazard rates from the marginal survival distributions be conditional on covariates. We use conditional copulas, first introduced by \cite{Patton2006b} who extended Sklar's Theorem \citep{Sklar1959} to account for the conditioning of the random variables on covariates.  \\

\noindent Let $T_1$ denote the time to the non-terminal event and $T_2$ denote the time to the terminal event with censoring time, $C$. The time to the first event is denoted by $X=\min(T_1,T_2,C)$ with event indicator $d_1=1$ if $T_1\leq \min(T_2,C)$ and $d_1=0$ otherwise. The time to the second event is denoted by $Y=\min(T_2,C)$ with event indicator $d_2=1$ if $T_2\leq C$ and $d_2=0$ otherwise. \\

\noindent We let the marginal hazard rates and the association parameter depend on covariates, $\mathbf{W} = (W_1,...,W_p)$. The marginal survival functions are given by $S_{T_1|\mathbf{W}}(t_1|\textbf{w}) = P(T_1>t_1|\textbf{w})$ and $S_{T_2|\mathbf{W}}(t_2|\textbf{w})= P(T_2>t_2|\textbf{w})$ for the non-terminal and terminal events, respectively. The marginal probability density functions (PDFs) are denoted by $f_{T_1|\mathbf{W}}(t_1|\textbf{w})$ and $f_{T_2|\mathbf{W}}(t_2|\textbf{w})$ for the non-terminal and terminal events, respectively. \\

\noindent We estimate the association between the survival probabilities of individuals, using univariate survival functions in the copula function. This changes the interpretation of the strength of association between the endpoints \citep{Renfro2015}, compared to the use of the cumulative distribution function (CDF) in \cite{Patton2006b}. We represent the joint survival function using the copula function, $C_\theta$, with association parameter $\theta$,
\begin{equation}\label{copula_c5}
S_B(t_1,t_2|\textbf{w})=P(T_1>t_1,T_2>t_2|\textbf{w})=C_\theta(S_{T_1|\mathbf{W}}(t_1|\textbf{w}),S_{T_2|\mathbf{W}}(t_2|\textbf{w})|\textbf{w}),
\end{equation}
where the subscript $B$ is used to indicate the bivariate nature of the function.

\noindent Then, the copula density function is as follows,

\begin{equation}
c_\theta(S_{T_1|\mathbf{W}}(t_1|\textbf{w}),S_{T_2|\mathbf{W}}(t_2|\textbf{w})) =\frac{\partial^2 C_\theta(S_{T_1|\mathbf{W}}(t_1|\textbf{w}),S_{T_2|\mathbf{W}}(t_2|\textbf{w})|\textbf{w})}{\partial S_{T_1|\mathbf{W}}(t_1|\textbf{w}) \partial S_{T_2|\mathbf{W}}(t_2|\textbf{w})}.
\end{equation}

\noindent The joint PDF can be expressed using the copula density function and the marginal PDFs,

\begin{equation}
f_{B}(t_1,t_2|\textbf{w})=c_\theta(S_{T_1|\mathbf{W}}(t_1|\textbf{w}),S_{T_2|\mathbf{W}}(t_2|\textbf{w})|\textbf{w})f_{T_1|\mathbf{W}}(t_1|\textbf{w})f_{T_2|\mathbf{W}}(t_2|\textbf{w}).
\end{equation}

\noindent We consider four different types of copula, the Normal, Clayton, Frank and Gumbel copulas, for which the definitions are given in \cite{Sorrell2022}. These copula functions cover a range of dependence structures and offer different shapes of the joint survival function in equation \eqref{copula_c5}. For each of these copulas, the association parameter $\theta$ is univariate. For the marginal distributions for both events, we consider the Exponential, Weibull and Gompertz distributions, which are commonly used parametric survival models. We use them to illustrate the methods and applications, however these can be easily extended to other parametric survival distributions.

\subsection{Likelihood}
\noindent Let $\mathbf{\Theta}$ be a vector of the parameters of the marginal distributions and the association parameter $\theta$ that are conditional on covariates. Then, the likelihood function is given as follows,

\begin{align} \label{lik5}
\begin{split}
L(\mathbf{\Theta})=&\prod^n_{i=1} \left( c_\theta(S_{T_{1}|\mathbf{W}}(t_{i,1}|\textbf{w}_i),S_{T_{2}|\mathbf{W}}(t_{i,2}|\textbf{w}_i)|\textbf{w}_i)f_{T_1|\mathbf{W}}(t_{i,1}|\textbf{w}_i)f_{T_2|\mathbf{W}}(t_{i,2}|\textbf{w}_i) \right)^{d_{i,1} d_{i,2}} \\
&\left (\frac{\partial C_\theta(S_{T_1|\mathbf{W}}(t_{i,1}|\textbf{w}_i),S_{T_2|\mathbf{W}}(t_{i,2}|\textbf{w}_i)|\textbf{w}_i)}{\partial S_{T_1|\mathbf{W}}(t_{i,1}|\textbf{w}_i)} f_{T_1|\mathbf{W}}(t_{i,1}|\textbf{w}_i)\right)^{d_{i,1}(1-d_{i,2})}\\
&\left(\frac{\partial C_\theta(S_{T_1|\mathbf{W}}(t_{i,1}|\textbf{w}_i),S_{T_2|\mathbf{W}}(t_{i,2}|\textbf{w}_i)|\textbf{w}_i) }{\partial S_{T_2|\mathbf{W}}(t_{i,2}|\textbf{w}_i)} f_{T_2|\mathbf{W}}(t_{i,2}|\textbf{w}_i) \right)^{(1-d_{i,1})d_{i,2}}\\
&\left(C_\theta(S_{T_1|\mathbf{W}}(t_{i,1}|\textbf{w}_i),S_{T_2|\mathbf{W}}(t_{i,2}|\textbf{w}_i)|\textbf{w}_i)\right)^{(1-d_{i,1})(1-d_{i,2})},
\end{split}
\end{align}


\noindent where $n$ is the total number of individuals.

\noindent Below, we give the formulae for the likelihood when the four types of copula, the Normal, Clayton, Frank and Gumbel, are used. Table \ref{linksfortheta} summarises the different link functions for each copula that are used to ensure the association parameters are within the permissible ranges within the respective copula functions.

\begin{table}[!h]
\begin{center}
\begin{tabular}{|l|l|l|}\hline
Copula $C_{\theta}$ & Link function for $\theta$ & Range of $\theta$ \\\hline
Normal & $\displaystyle\theta = \frac{\exp(2 (b_0 + b_1 W_1 + ... + b_p W_p))-1}{\exp(2 (b_0 + b_1 W_1 + ... + b_p W_p))+1}$ & $\theta \in[-1,1]$\\[0.3cm]
Clayton & $\displaystyle \theta = \exp(b_0 + b_1 W_1 + ... + b_p W_p)$ & $\theta \in (0,\infty)$\\[0.3cm]
Frank & $\displaystyle\theta = b_0 + b_1 W_1 + ... + b_p W_p$ & $\theta \in (-\infty, \infty)$\\[0.3cm]
Gumbel & $\displaystyle\theta = \exp(b_0 + b_1 W_1 + ... + b_p W_p)+1$ & $\theta \in [1,\infty)$ \\\hline
\end{tabular}
\end{center}
\caption{Link functions for the association parameter $\theta$.}
\label{linksfortheta}
\end{table}

\noindent For the Normal copula, the association parameter $\theta = \rho$, where $\rho$ is the Pearson correlation coefficient, and the likelihood function of equation \eqref{lik5} is given by
\begin{align}
\begin{split}
& L_N(\mathbf{\Theta})=\prod_{i=1}^n \left(\frac{1}{\sqrt{1-\rho^2}} \exp\left(\frac{1}{2(1-\rho^2)} \right.\right.\\
&\left.\left. 2\rho\Phi^{-1}(S_{T_1|\mathbf{W}}(t_{i,1}|\mathbf{w}_i))\Phi^{-1}(S_{T_2|\mathbf{W}}(t_{i,2}|\mathbf{w}_i))-\rho^2(\Phi^{-1}(S_{T_1|\mathbf{W}}(t_{i,1}|\mathbf{w}_i))^2 \right.\right.\\
&\left.\left.+\Phi^{-1}(S_{T_2|\mathbf{W}}(t_{i,2}|\mathbf{w}_i))^2)\right) f_{T_1|\mathbf{W}}(t_{i,1}|\mathbf{w}_i)f_{T_2|\mathbf{W}}(t_{i,2}|\mathbf{w}_i) \right)^{d_{i,1} d_{i,2}} \\
&\left (\Phi\left(\frac{\Phi^{-1}(S_{T_2|\mathbf{W}}(t_{i,2}|\mathbf{w}_i))-\rho\Phi^{-1}(S_{T_1|\mathbf{W}}(t_{i,1}|\mathbf{w}_i)))}{\sqrt{1-\rho^2}}\right) f_{T_1|\mathbf{W}}(t_{i,1}|\mathbf{w}_i)\right)^{d_{i,1}(1-d_{i,2})}\\
&\left(\Phi\left(\frac{\Phi^{-1}(S_{T_1|\mathbf{W}}(t_{i,1}|\mathbf{w}_i))-\rho\Phi^{-1}(S_{T_2|\mathbf{W}}(t_{i,2}|\mathbf{w}_i))}{\sqrt{1-\rho^2}}\right) f_{T_2|\mathbf{W}}(t_{i,2}|\mathbf{w}_i) \right)^{(1-d_{i,1})d_{i,2}}\\
&\left(C_{\rho} (S_{T_1|\mathbf{W}}(t_{i,1}|\mathbf{w}_i),S_{T_2|\mathbf{W}}(t_{i,2}|\mathbf{w}_i))\right)^{(1-d_{i,1})(1-d_{i,2})},
\end{split}
\end{align}\label{c5_normal}

\noindent where $\Phi$ is the CDF of the standard normal distribution.

\normalsize

\noindent For the Clayton copula, the likelihood function is given by

$$
L_C(\mathbf{\Theta})=\prod_{i=1}^n \left((1+\theta)\frac{(C_\theta (S_{T_1|\mathbf{W}}(t_{i,1}|\mathbf{w}_i),S_{T_2|\mathbf{W}}(t_{i,2}|\mathbf{w}_i)|\mathbf{w}_i))^{1+2\theta}}{(S_{T_1|\mathbf{W}}(t_{i,1}|\mathbf{w}_i)S_{T_2|\mathbf{W}}(t_{i,2}|\mathbf{w}_i))^{1+\theta}}\right.
$$
$$
\left.f_{T_1|\mathbf{W}}(t_{i,1}|\mathbf{w}_i)f_{T_2|\mathbf{W}}(t_{i,2}|\mathbf{w}_i) \right)^{d_{i,1} d_{i,2}}$$
$$ \left (\left(\frac{ C_\theta (S_{T_1|\mathbf{W}}(t_{i,1}|\mathbf{w}_i),S_{T_2|\mathbf{W}}(t_{i,2}|\mathbf{w}_i)|\mathbf{w}_i)}{S_{T_1|\mathbf{W}}(t_{i,1}|\mathbf{w}_i)}\right)^{1+\theta} f_{T_1|\mathbf{W}}(t_{i,1}|\mathbf{w}_i) \right)^{d_{i,1}(1-d_{i,2})}$$
$$ \left(\left(\frac{ C_\theta (S_{T_1|\mathbf{W}}(t_{i,1}|\mathbf{w}_i),S_{T_2|\mathbf{W}}(t_{i,2}|\mathbf{w}_i)|\mathbf{w}_i)}{S_{T_2|\mathbf{W}}(t_{i,2}|\mathbf{w}_i)}\right)^{1+\theta} f_{T_2|\mathbf{W}}(t_{i,2}|\mathbf{w}_i) \right)^{(1-d_{i,1})d_{i,2}} $$
\begin{equation}\left(C_{\theta} (S_{T_1|\mathbf{W}}(t_{i,1}|\mathbf{w}_i),S_{T_2|\mathbf{W}}(t_{i,2}|\mathbf{w}_i))\right)^{(1-d_{i,1})(1-d_{i,2})}. \label{claylik}
\end{equation}\label{c5_clayton}

\noindent  For the Frank copula, the likelihood function is given by

\begin{align}
\begin{split}
L_F(\mathbf{\Theta})=&\prod_{i=1}^n \left(\frac{\theta e^{\theta C_{\theta}(S_{T_1|\mathbf{W}}(t_{i,1}|\mathbf{w}_i),S_{T_2|\mathbf{W}}(t_{i,2}|\mathbf{w}_i)|\mathbf{w}_i)}(e^{\theta C_{\theta}(S_{T_1|\mathbf{W}}(t_{i,1}|\mathbf{w}_i),S_{T_2|\mathbf{W}}(t_{i,2}|\mathbf{w}_i)|\mathbf{w}_i)}-1)}{(e^{\theta S_{T_1|\mathbf{W}}(t_{i,1}|\mathbf{w}_i)}-1)(e^{\theta S_{T_2|\mathbf{W}}(t_{i,2}|\mathbf{w}_i)}-1)} \right. \\
&\left. f_{T_1|\mathbf{W}}(t_{i,1}|\mathbf{w}_i)f_{T_2|\mathbf{W}}(t_{i,2}|\mathbf{w}_i) \right)^{d_{i,1} d_{i,2}}\\
&\left (\frac{1-e^{\theta C_{\theta}(S_{T_1|\mathbf{W}}(t_{i,1}|\mathbf{w}_i),S_{T_2|\mathbf{W}}(t_{i,2}|\mathbf{w}_i)|\mathbf{w}_i)}}{1-e^{\theta S_{T_1|\mathbf{W}}(t_{i,1}|\mathbf{w}_i)}} f_{T_1|\mathbf{W}}(t_{i,1}|\mathbf{w}_i)\right)^{d_{i,1}(1-d_{i,2})}\\
&\left(\frac{1-e^{\theta C_{\theta}(S_{T_2|\mathbf{W}}(t_{i,2}|\mathbf{w}_i),S_{T_2|\mathbf{W}}(t_{i,2}|\mathbf{w}_i)|\mathbf{w}_i)}}{1-e^{\theta S_{T_2|\mathbf{W}}(t_{i,2}|\mathbf{w}_i)}} f_{T_2|\mathbf{W}}(t_{i,2}|\mathbf{w}_i) \right)^{(1-d_{i,1})d_{i,2}}\\
&\left(C_\theta(S_{T_1|\mathbf{W}}(t_{i,1}|\mathbf{w}_i),S_{T_2|\mathbf{W}}(t_{i,2}|\mathbf{w}_i)|\mathbf{w}_i)\right)^{(1-d_{i,1})(1-d_{i,2})}.
\end{split}
\end{align}\label{c5_frank}

\noindent Finally, for the Gumbel copula the likelihood function is given by

\begin{align}
\begin{split}
& L_G(\mathbf{\Theta})=\prod_{i=1}^n \left(\frac{1}{S_{T_1|\mathbf{W}}(t_{i,1}|\mathbf{w}_i)S_{T_2|\mathbf{W}}(t_{i,2}|\mathbf{w}_i)(-\log(C_{\theta}(S_{T_1|\mathbf{W}}(t_{i,1}|\mathbf{w}_i),S_{T_2|\mathbf{W}}(t_{i,2}|\mathbf{w}_i)|w_i)))^{2\theta-1}}\right.\\
& \left. C_{\theta}(S_{T_1|\mathbf{W}}(t_{i,1}|\mathbf{w}_i),S_{T_2|\mathbf{W}}(t_{i,2}|\mathbf{w}_i)|\mathbf{w}_i)(-\log(S_{T_1|\mathbf{W}}(t_{i,1}|\mathbf{w}_i)))^{\theta-1}(-\log(S_{T_2|\mathbf{W}}(t_{i,2}|\mathbf{w}_i)))^{\theta-1}\right.\\
&\left. (\theta-1-\log(C_{\theta}(S_{T_1|\mathbf{W}}(t_{i,1}|\mathbf{w}_i),S_{T_2|\mathbf{W}}(t_{i,2}|\mathbf{w}_i)|\mathbf{w}_i)))f_{T_1|\mathbf{W}}(t_{i,1}|\mathbf{w}_i)f_{T_2|\mathbf{W}}(t_{i,2}|\mathbf{w}_i)\right)^{d_{i,1} d_{i,2}} \\
& \left (\frac{C_{\theta}(S_{T_1|\mathbf{W}}(t_{i,1}|\mathbf{w}_i),S_{T_2|\mathbf{W}}(t_{i,2}|\mathbf{w}_i)|\mathbf{w}_i)(-\log(S_{T_1|\mathbf{W}}(t_{i,1}|\mathbf{w}_i)))^{\theta-1}}{S_{T_1|\mathbf{W}}(t_{i,1}|\mathbf{w}_i)(-\log(C_{\theta}(S_{T_1|\mathbf{W}}(t_{i,1}|\mathbf{w}_i),S_{T_2|\mathbf{W}}(t_{i,2}|\mathbf{w}_i)|\mathbf{w}_i))^{\theta-1}} f_{T_1|\mathbf{W}}(t_{i,1}|\mathbf{w}_i)\right)^{d_{i,1}(1-d_{i,2})}\\
& \left(\frac{C_{\theta}(S_{T_2|\mathbf{W}}(t_{i,2}|\mathbf{w}_i),S_{T_2|\mathbf{W}}(t_{i,2}|\mathbf{w}_i)|\mathbf{w}_i)(-\log(S_{T_2|\mathbf{W}}(t_{i,2}|\mathbf{w}_i)))^{\theta-1}}{S_{T_2|\mathbf{W}}(t_{i,2}|\mathbf{w}_i)(-\log(C_{\theta}(S_{T_2|\mathbf{W}}(t_{i,2}|\mathbf{w}_i),S_{T_2|\mathbf{W}}(t_{i,2}|\mathbf{w}_i)|\mathbf{w}_i))^{\theta-1}}f_{T_2|\mathbf{W}}(t_{i,2}|\mathbf{w}_i) \right)^{(1-d_{i,1})d_{i,2}}\\
& \left(C_{\theta} (S_{T_1|\mathbf{W}}(t_{i,1}|\mathbf{w}_i),S_{T_2|\mathbf{W}}(t_{i,2}|\mathbf{w}_i))\right)^{(1-d_{i,1})(1-d_{i,2})}.
\label{glikt}
\end{split}
\end{align}\label{c5_gumbel}

\noindent For the Normal, Clayton and Gumbel copulas, the variance of the association parameter $\theta$ can be approximated using the Delta method. We describe for the case of one covariate as well as for several covariates in the Supplementary Materials, in Section 1.2 for the Normal, Section 1.3 for the Clayton and Section 1.4 for the Gumbel copula, respectively.

\subsection{Event times models}\label{margins}

We consider the Exponential, Weibull and Gompertz distributions for the marginal distributions of event times. For each model, we specify the link function for marginal parameters. An Exponential model assumes a constant hazard rate, while the Weibull and Gompertz models allow for increasing, constant or decreasing hazard rates. The methods can be easily extended to other parametric survival distributions such as a Log-normal or Log-logistic distributions. \\

\textbf{Exponential event times}

\noindent Assume that the marginal distributions for both events follow the Exponential distribution with hazard rates $\lambda_1$ and $\lambda_2$ for the non-terminal and terminal events, respectively. Then, the survival functions are given by $S_{T_1|\textbf{W}}(t_1)=\exp(-\lambda_1 t_1)$ and $S_{T_2|\textbf{W}}(t_2)=\exp(-\lambda_2 t_2)$ for the non-terminal and terminal events, respectively. We incorporate covariates $W_1,\ldots,W_p$ into the hazard rates of both events:

\begin{equation}\label{c5_l1_multi}
\lambda_1=\exp(a_0+a_1  W_1 + ... + a_p W_p),
\end{equation}
\begin{equation}\label{c5_l2_multi}
\lambda_2=\exp(c_0+c_1  W_1 + ...+ c_p W_p),
\end{equation}

\noindent where $a_0,...,a_p$ are regression coefficients for the non-terminal event and $c_0,...,c_p$ are regression coefficients for the terminal event.
Therefore, the hazard ratios for a binary covariate $W_k$, where $k\in\{1,\ldots,p\}$, are given by

\begin{equation}\label{c5_hr1}\mbox{HR}_{NT}=\frac{\exp(a_0+a_k)}{\exp(a_0)}=\exp(a_k),
\end{equation}
\begin{equation}\label{c5_hr2}
\mbox{HR}_{T}=\frac{\exp(c_0+c_k)}{\exp(c_0)}=\exp(c_k),
\end{equation}

\noindent for the non-terminal and terminal events, respectively. \\

\textbf{Weibull event times}

Assume that the marginal distributions for both events follow the Weibull distributions with the PDFs

$$f_{T_j|\textbf{W}}(t_j) = \beta_j \alpha_j t^{\alpha_j -1}_j \exp(-\beta_j t^\alpha_j),$$

\noindent for $j=1,2$, where $\alpha_j$ is the shape parameter and $\beta_j$ is the scale parameter and $t_j$ represents the event time for the non-terminal and terminal events, respectively.

\noindent The survival function and hazard function are given by the following

$$S_{T_j|\textbf{W}}(t_j) = \exp(-\beta_j t^\alpha_j), $$
$$h_{T_j|\textbf{W}}(t_j) = \beta_j \alpha_j t^{\alpha_j-1}_j, $$

\noindent for $j=1,2$.

\noindent We consider the case where the shape parameters $\alpha_1$ and $\alpha_2$ are constant and the scale parameters  $\beta_1$ and $\beta_2$ depend on the covariates $\textbf{W} = (W_1,\ldots,W_p)$  using the following calibration functions:

$$\beta_1=\exp(a_0+a_1  W_1 + ... + a_p W_p),$$
$$\beta_2=\exp(c_0+c_1  W_1 + ... + c_p W_p),$$

\noindent where $a_0,...,a_p$ are regression coefficients for the non-terminal event and $c_0,...,c_p$ are regression coefficients for the terminal event.

\noindent Therefore, the hazard ratios for a binary covariate $W_k$, where $k\in\{1,\ldots,p\}$, are given by

$$ \mbox{HR}_{NT}=\frac{\exp(a_0+a_k) \alpha_1 t^{\alpha_1 -1}}{\exp(a_0)\alpha_1 t^{\alpha_1 -1}}=\exp(a_k),$$
$$\mbox{HR}_{T}=\frac{\exp(c_0+c_k) \alpha_2 t^{\alpha_2 -1}}{\exp(c_0)\alpha_2 t^{\alpha_2 -1}}=\exp(c_k),$$

\noindent respectively. These are in the same form as for the Exponential event times described  in equations (\ref{c5_hr1}) and (\ref{c5_hr2}). \\

\textbf{Gompertz event times}

Assume that the marginal distributions for both events follow the Gompertz distributions with the PDFs

$$f_{T_j|\textbf{W}}(t_j) = \lambda_j \exp\left(\gamma_j t_j - \frac{\lambda_j}{\gamma_j}(\exp(\gamma_j t_j)-1)\right),$$

\noindent for $j=1,2$, where $\gamma_j$ is the shape parameter and $\lambda_j$ is the rate parameter and $t_j$ represents the event time for the non-terminal and terminal events, respectively.

\noindent The survival function and hazard function are given by the following

$$S_{T_j|\textbf{W}}(t_j) = \exp\left(-\frac{\lambda_j}{\gamma_j}(\exp(\gamma_j t_j) -1)\right), $$
$$h_{T_j|\textbf{W}}(t_j) = \lambda_j \exp(\gamma_j t_j),$$

\noindent for $j=1,2$.

\noindent We consider the case where the shape parameters $\gamma_1$ and $\gamma_2$ are constant and the rate parameters  $\lambda_1$ and $\lambda_2$ depend on the covariates $W_1,\ldots,W_p$ using the following calibration functions:

$$\lambda_1=\exp(a_0+a_1  W_1 + ... + a_p W_p),$$
$$\lambda_2=\exp(c_0+c_1  W_1 + ... + c_p W_p),$$

\noindent where $a_0,...,a_p$ are regression coefficients for the non-terminal event and $c_0,...,c_p$ are regression coefficients for the terminal event.

\noindent  Therefore, the hazard ratios for a binary covariate $W_k$, where $k\in\{1,\ldots,p\}$, are

$$ \mbox{HR}_{NT}=\frac{\exp(a_0+a_k) \exp(\gamma_1 t)}{\exp(a_0)\exp(\gamma_1 t)}=\exp(a_k),$$
$$\mbox{HR}_{T}=\frac{\exp(c_0+c_k) \exp(\gamma_2 t)}{\exp(c_0)\exp(\gamma_2 t)}=\exp(c_k),$$

\noindent respectively. These are in the same form as for the Exponential and Weibull event times.

For all three marginal models, the variances of the hazard ratios can be approximated by

\begin{equation}\label{var_hr_nt}\mbox{Var}(\mbox{HR}_{NT})=\exp(2\hat{a}_k)\mbox{Var}(\hat{a}_k), \end{equation}
\begin{equation}\label{var_hr_t} \mbox{Var}(\mbox{HR}_{T})=\exp(2\hat{c}_k)\mbox{Var}(\hat{c}_k), \end{equation}

\noindent using the Delta method described in Section 1.1 in the Supplementary Materials.

\subsection{Estimation}

\noindent The regression coefficients for the marginal distributions and the association parameters are estimated by maximising the log-likelihood, i.e. the logarithm of (\ref{lik5}). This is achieved by using the function \texttt{optim} in package \texttt{R} \citep{R} in the practical application presented in this paper. The 95\% confidence intervals (CI) are constructed using the Fisher Information matrix.


\section{Application} \label{sec:analysis}

\noindent We apply the methods described in Section \ref{sec:meth} to the UKTR data set introduced in Section \ref{sec:data}. The data set contains time to graft failures and time to death of kidney transplant recipients. To illustrate the use of the methods, we include the following binary covariates, recipient age group ($>50$ years and $\le 50$ years), recipient sex (female and male) and donor type (living and deceased donors), in the analysis of both survival endpoints. We allow both the hazard rates and the association between the survival endpoints to vary with these covariates. We estimate the hazard ratios of the non-terminal and terminal events along with the association parameters for the reference and covariate groups.

We use Exponential, Weibull and Gompertz distributions to model the survival time. We apply four different copula functions to describe the association between the survival endpoints and we select the best fitting models using the Akaike Information Criterion, AIC \citep{Akaike1973}. The estimated hazard ratios for each covariate and the regression coefficients of the covariates for the association parameter are provided in Tables \ref{table:hr_allcovs}, \ref{table:hr_allcovs_Gompertz} and \ref{table:hr_allcovs_Weibull} for the Exponential, Gompertz and Weibull survival distributions, respectively. The results for the Cox model are presented in Table \ref{table:hr_allcovs}. Moreover, in Tables \ref{table:hr_allcovs}-\ref{table:hr_allcovs_Weibull} the computational time for each model is reported for a computer with processor $11^{th}$ Gen Intel (R) Core(TM) i7-1165G7 CPU $@$ 2.80GHz.

\subsection{Hazard ratios}

\subsubsection{Graft failure following transplant}
Across all considered models, sex is not found to be associated with the risks for graft failure. Compared to a deceased donor transplant, living donor transplant is associated with lower risk for graft failure. Older age ($>50$ years) is found to be associated with increased risk across all considered copula models except for two cases: Normal and Gumbel copulas with Weibull survival model, where there is no association found. This is in contrast with the Cox model where the older age group is found to be associated with lower risk of graft failure (HR: 0.911, 95\% CI: 0.872 to 0.952, Table \ref{table:hr_allcovs}). This may be in part due to the censoring of graft failure by death in the Cox model, the older individuals who die before experiencing graft failure may be seen as less likely to experience graft failure.

In the fitted Cox model for graft failure, there was evidence for non-proportional hazards for all three covariates. Using Grambsch and Therneau's approach to diagnose non-proportionality, we obtained P-values of 0.091, 0.017  and $<$0.001 for sex, age and donor type, respectively.

\subsubsection{Death following transplant}
For all considered models, including the Cox model and all copula models, all three covariates: sex, age and donor type, are found to be associated with death after transplant. In particular, female sex,  living donor transplant and younger age ($\le50$ years) are associated with lower risk of death. In contrast, male sex, deceased donor transplant and older age ($>50$ years) are associated with higher risk of death. These findings are consistent across all models and the estimated hazard ratios are fairly similar.

In the fitted Cox model for death, there was evidence for non-proportional hazards for donor type (P$<$0.001), whilst there was insufficient evidence for non-proportional hazards for sex (P=0.936) and age (P=0.882).

\subsection{Association between graft failure and death}

\noindent The results from all the copula survival models showed that the association between graft failure and death is stronger for individuals in the older age group compared to the younger age group.

In most of the fitted copula models, we observe no difference between female and male recipients for the association between graft failure and death. The exceptions are the Clayton copula models and Frank copula Gompertz model which show stronger association between these two end points for female recipients. Similarly, in most of the fitted copula models, we observe no difference between living and deceased donors for the association between graft failure and death. The exceptions are the Clayton copula models and Frank copula Exponential model which show stronger association between these two end points for living donor recipients.

\subsection{Results for the preferred model}

\noindent
In our real data analyses, we reported the AIC value for each model. However, we did not compare Cox model with the copula models using AIC. This is because the Cox model was fitted to graft failure and death, separately, and the Cox model has a partial likelihood instead of a full likelihood. In contrast, the copula survival model analysed both graft failure and death jointly, and has a full likelihood. Hence, the AIC values for the Cox model where outcomes were analysed separately and the copula-based parametric survival models where outcomes were analysed jointly are not comparable. However, we used the AIC to compare between the copula survival models. For each survival distribution,  the Frank copula model is the preferred according to the AIC criterion (Tables \ref{table:hr_allcovs}, \ref{table:hr_allcovs_Gompertz}, \ref{table:hr_allcovs_Weibull}). Moreover, for each copula model, the Weibull survival distribution provides the lowest AIC.

\noindent
In the Frank copula Weibull survival model, the association between graft failure and death is affected by age, with individuals in the older age group having a stronger association compared to those in the younger age group. The hazard ratio of graft failure for living donors is 0.560 (95\% CI: 0.532 to 0.588), and for death is 0.519 (95\% CI: 0.490 to 0.549), indicating the living donor recipient group is at lower risk for both events. Female sex is associated with lower risk for death compared to men, with hazard ratio 0.920 (95\% CI: 0.882 to 0.957). The hazard ratio of graft failure for the older age group is 1.202 (95\% CI: 1.152 to 1.251) and for death is 3.734 (95\% CI: 3.565 to 3.903), indicating the older age group is at higher risk of graft failure and death.

\begin{figure*}[h!t!]
    \centering
    \begin{subfigure}[b]{0.5\textwidth}
        \centering
        \includegraphics[width=2.8in]{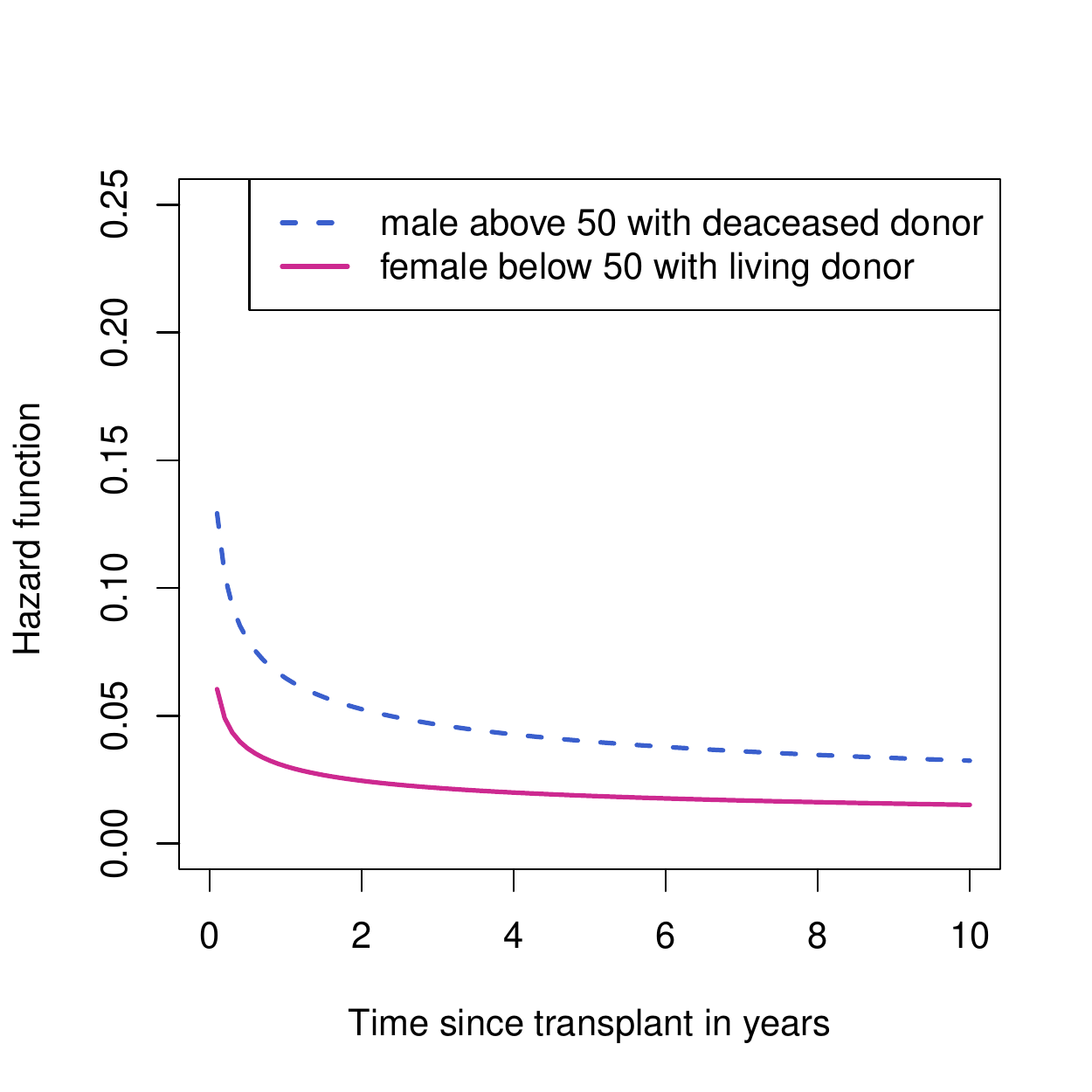}
        \caption{\label{nt} Graft failure}
    \end{subfigure}%
    \begin{subfigure}[b]{0.5\textwidth}
        \centering
        \includegraphics[width=2.8in]{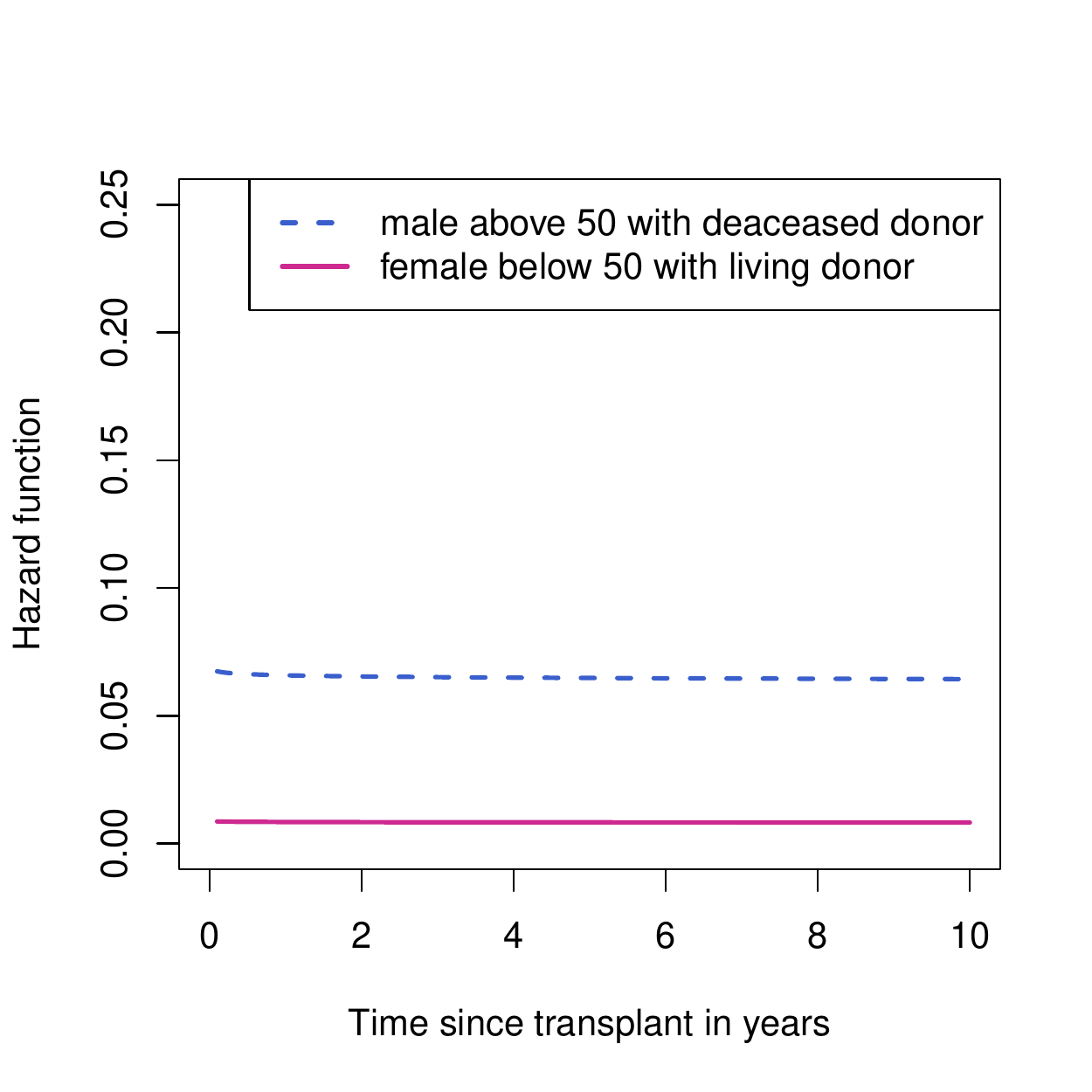}
        \caption{\label{terminal} Death}
    \end{subfigure}

 \caption{Estimated hazard functions for two subgroups of patients: male recipients above 50 years with deceased donor and female recipients aged below 50 years with living donor, using Frank copula with Weibull survival times.}\label{hazard}
\end{figure*}

The Frank copula Weibull survival model suggests that in general male patients above 50 years who received a transplant from a deceased donor are at highest overall risk for death following kidney transplant. At the other extreme, younger female patients who received a transplant from a living donor are at the lowest overall risk for death. The hazard functions estimated from the preferred model are presented in Figure \ref{hazard} for these two subgroups, respectively. For both subgroups, the hazard for graft failure is greatest immediately following the transplant and gradually decreases within 4 years where it is stabilised (Figure \ref{nt}). The hazard for death is stable over time since transplant (Figure \ref{terminal}). For male recipients aged $>50$ years with a deceased donor, for the first 2 years following kidney transplant, the hazard  for graft failure is greater than for death and after that the order is reversed.

\begin{sidewaystable}[ht!]
\centering
\begin{tabular}{| l  lll rr|}
\hline
Model/ Covariate & \multicolumn{2}{c}{Hazard Ratio (95\%CI)}  & Regression coefficients on & AIC& Time  \\

 &  Graft Failure (GF) & Death (D) &  association parameter& & (minutes) \\
\hline
\multicolumn{3}{|l}{Cox model}   & &174,873.6 (GF)& 0.004\\
 &  && & 153,814.9 (D) & 0.005\\
 \phantom{xxx} Age group: $>50$  & 0.911 (0.872, 0.952) & 3.926 (3.745, 4.116)& NA &&\\
 \phantom{xxx} Sex: Female       & 1.029 (0.986, 1.073) & 0.881 (0.841, 0.922) & NA& & \\
 \phantom{xxx} Donor type: Living& 0.615 (0.584, 0.648) & 0.539 (0.506, 0.574) & NA & &\\
\hline\multicolumn{3}{|l}{Normal copula Exponential survival model} & &										147545.5&	123.6\\
\phantom{xxx} Age group: $>50$      &	1.121	 (1.074, 	1.169) &	3.748	 (3.576, 	3.920)  &	0.285	 (0.248, 	0.321) &	&	\\
\phantom{xxx} Sex: Female           &	1.015	 (0.973, 	1.058) &	0.900	 (0.860, 	0.939) &	0.021	 ($-$0.015, 	0.058) &	&	\\
\phantom{xxx} Donor type: Living    &	0.600	 (0.569, 	0.630) &	0.522	 (0.490, 	0.553) &	0.027	 ($-$0.025, 	0.078) &	&	\\
\hline\multicolumn{3}{|l}{Clayton copula Exponential survival model} & &										146956.7&	1.63\\
\phantom{xxx} Age group: $>50$      &	1.383	 (1.329, 	1.437) &	3.863	 (3.691, 	4.036) &	1.093	 (0.983, 	1.203) &	&	\\
\phantom{xxx} Sex: Female           &	1.006	 (0.967, 	1.045) &	0.934	 (0.898, 	0.969) &	0.139	 (0.039, 	0.239) &	&	\\
\phantom{xxx} Donor type: Living    &	0.587	 (0.559, 	0.616) &	0.540	 (0.512, 	0.568) &	0.529	 (0.384, 	0.674) &	&	\\
\hline\multicolumn{3}{|l}{Frank copula Exponential survival model} & &										146660.5&	2.19\\
\phantom{xxx} Age group: $>50$      &	1.361	 (1.309, 	1.414) &	3.861	 (3.689, 	4.032) &	5.068	 (4.600, 	5.537) &	&	\\
\phantom{xxx} Sex: Female           &	1.00	 (0.961, 	1.039) &	0.930	 (0.893, 	0.966) &	0.352	 ($-$0.059, 	0.762) &	&	\\
\phantom{xxx} Donor type: Living    &	0.586	 (0.557, 	0.615) &	0.535	 (0.506, 	0.564) &	0.861	 (0.211, 	1.511) &	&	\\
\hline\multicolumn{3}{|l}{Gumbel copula Exponential survival model} & &										147827.8&	4.29\\
\phantom{xxx} Age group: $>50$      &	1.134	 (1.086, 	1.182) &	3.667	 (3.498, 	3.837) &	1.355	 (1.204, 	1.507) &	&	\\
\phantom{xxx} Sex: Female           &	1.015	 (0.973, 	1.057) &	0.899	 (0.859, 	0.939) &	0.056	 ($-$0.079, 	0.191) &	&	\\
\phantom{xxx} Donor type: Living    &	0.603	 (0.573, 	0.634) &	0.530	 (0.498, 	0.562) &	$-$0.039	 ($-$0.231, 	0.153) &	&	\\
\hline
\end{tabular}
\caption{Cox model and Exponential survival copula models: Results from analysing the UK Transplant Registry. The binary covariates are defined as follows, sex: female vs male (reference), donor type: living vs deceased (reference), and age group: $>50$ years vs $\leq 50$ (reference).}
\label{table:hr_allcovs}
\end{sidewaystable}

\begin{sidewaystable}[ht!]
\centering
\begin{tabular}{| l  c c c r r|}
\hline
Model/ Covariate & \multicolumn{2}{c}{Hazard Ratio (95\%CI)}  &Regression coefficients on  & AIC & Time\\

 &  Graft failure & Death& association prameter & & (minutes) \\

\hline\multicolumn{3}{|l}{Normal copula Gompertz survival model} & &										146872.6&	250.8\\
\phantom{xxx} Age group: $>50$          &	1.080	 (1.033, 	1.126) &	4.163	 (3.969, 	4.357) &	0.282	 (0.246, 	0.318) &	&	\\
\phantom{xxx} Sex: Female               &	1.018	 (0.976, 	1.061) &	0.893	 (0.854, 	0.932) &	0.024	 ($-$0.012, 	0.060) &	&	\\
\phantom{xxx} Donor type: Living        &	0.594	 (0.563, 	0.624) &	0.553	 (0.520, 	0.586) &	0.020	 ($-$0.03, 	0.071) &	&	\\
\hline\multicolumn{3}{|l}{Clayton copula Gompertz survival model} & &										146708.3&	5.28\\
\phantom{xxx} Age group: $>50$          &	1.296	 (1.242, 	1.349) &	3.935	 (3.756, 	4.114) &	0.947	 (0.835, 	1.059) &	&	\\
\phantom{xxx} Sex: Female               &	0.999	 (0.959, 	1.039) &	0.909	 (0.873, 	0.945) &	0.161	 (0.059, 	0.263) &	&	\\
\phantom{xxx} Donor type: Living        &	0.592	 (0.562, 	0.622) &	0.555	 (0.524, 	0.586) &	0.287	 (0.118, 	0.456) &	&	\\
\hline\multicolumn{3}{|l}{Frank copula Gompertz survival model} & &										146409.4&	4.81\\
\phantom{xxx} Age group: $>50$          &	1.232	 (1.182, 	1.283) &	3.949	 (3.771, 	4.127) &	2.070	 (1.655, 	2.486) &	&	\\
\phantom{xxx} Sex: Female               &	1.003	 (0.963, 	1.042) &	0.919	 (0.881, 	0.957) &	0.639	 (0.233, 	1.045) &	&	\\
\phantom{xxx} Donor type: Living        &	0.544	 (0.516, 	0.572) &	0.498	 (0.469, 	0.528) &	0.423	 ($-$0.240, 	1.087) &	&	\\
\hline\multicolumn{3}{|l}{Gumbel copula Gompertz survival model} & &										147036.8&	6.67\\
\phantom{xxx} Age group: $>50$          &	1.065	 (1.020, 	1.111) &	4.11	 (3.918, 	4.303) &	1.401	 (1.251, 	1.505) &	&	\\
\phantom{xxx} Sex: Female               &	1.012	 (0.970, 	1.054) &	0.891	 (0.852, 	0.931) &	$-$0.017	 ($-$0.152, 	0.118) &	&	\\
\phantom{xxx} Donor type: Living        &	0.596	 (0.566, 	0.627) &	0.557	 (0.524, 	0.591) &	$-$0.096	 ($-$0.29, 	0.097) &	&	\\

\hline
\end{tabular}
\caption{Gompertz survival copula models: Results from analysing the UK Transplant Registry. The binary covariates are defined as follows, sex: female vs male (reference), donor type: living vs deceased (reference), and age group: $>50$ years vs $\leq 50$ (reference).}
\label{table:hr_allcovs_Gompertz}
\end{sidewaystable}

\begin{sidewaystable}[ht!]
\centering
\begin{tabular}{| l  lll rr|}
\hline
Model/ Covariate & \multicolumn{2}{c}{Hazard Ratio (95\%CI)}  & Regression coefficients  & AIC& Time\\

 &  Graft failure & Death& on association parameter &  &(minutes)\\

\hline\multicolumn{3}{|l}{Normal copula Weibull survival model} & &										145038.4&	201.0\\
\phantom{xxx} Age group: $>50$      &	1.005	 (0.962, 	1.048) &	3.692	 (3.523, 	3.860)  &	0.283	 (0.240, 	0.326) &	&	\\
\phantom{xxx} Sex: Female           &	1.022	 (0.979, 	1.064) &	0.904	 (0.865, 	0.943) &	0.028	 ($-$0.014, 	0.071) &	&	\\
\phantom{xxx} Donor type: Living    &	0.573	 (0.544, 	0.602) &	0.520	 (0.489, 	0.551) &	0.034	 ($-$0.027, 	0.094) &	&	\\
\hline\multicolumn{3}{|l}{Clayton copula Weibull survival model} & &										145137.8&	8.03\\
\phantom{xxx} Age group: $>50$      &	1.212	 (1.160, 	1.264) &	3.699	 (3.532, 	3.867) &	0.907	 (0.782, 	1.032) &	&	\\
\phantom{xxx} Sex: Female           &	1.024	 (0.983, 	1.066) &	0.924	 (0.887, 	0.96)  &	0.162	 (0.054, 	0.271) &	&	\\
\phantom{xxx} Donor type: Living    &	0.562	 (0.533, 	0.590)  &	0.527	 (0.498, 	0.556) &	0.479	 (0.314, 	0.643) &	&	\\
\hline\multicolumn{3}{|l}{Frank copula Weibull survival model} & &										144704.7&	6.16\\
\phantom{xxx} Age group: $>50$      &	1.202	 (1.152, 	1.251) &	3.734	 (3.565, 	3.903) &	3.954	 (3.488, 	4.402) &	&	\\
\phantom{xxx} Sex: Female           &	1.005	 (0.965, 	1.045) &	0.920	 (0.882, 	0.957) &	0.171	 ($-$0.251, 	0.592) &	&	\\
\phantom{xxx} Donor type: Living    &	0.560	 (0.532, 	0.588) &	0.519	 (0.490, 	0.549) &	0.493	 ($-$0.174, 	1.160) &	&	\\
\hline\multicolumn{3}{|l}{Gumbel copula Weibull survival model} & &										145276.1&	6.27\\
\phantom{xxx} Age group: $>50$      &	0.992	 (0.949, 	1.034) &	3.600	 (3.436, 	3.765) &	1.073	 (0.935, 	1.211) &	&	\\
\phantom{xxx} Sex: Female           &	1.020	 (0.978, 	1.062) &	0.906	 (0.867, 	0.945) &	0.044	 ($-$0.08, 	0.169) &	&	\\
\phantom{xxx} Donor type: Living    &	0.581	 (0.552, 	0.611) &	0.520	 (0.489, 	0.551) &	$-$0.043	 ($-$0.218, 	0.132) &	&	\\

\hline
\end{tabular}
\caption{Weibull survival copula models: results from analysing the UK Transplant Registry. The binary covariates are defined as follows, sex: female vs male (reference), donor type: living vs deceased (reference), and age group: $>50$ years vs $\leq 50$ (reference).}
\label{table:hr_allcovs_Weibull}
\end{sidewaystable}

\section{Simulation study} \label{sec:sim}

We conduct two simulation studies. In simulation study 1, we aim to assess the performance of the proposed bivariate copula regression models for semi-competing risk data. We simulate data to mimic the real example of renal transplant data. We compare the performance of three models: the Cox proportional hazards model; the bivariate copula regression models with predictors for the hazard rates (Copula model 1); and the bivariate copula regression models with predictors for both hazard rates and the association parameters (Copula model 2).

In simulation study 2, we aim to assess the effect of misspecification of the survival distributions on the estimation of the model parameters. We use the AIC value to select the best fitting model for the simulated data  and calculate the percentage that the underlying survival model is correctly selected.

\subsection{Design}
\noindent We simulate data with known marginal hazard rates for the non-terminal and terminal events,  and the correlation between the two event times. We assess the accuracy of the estimates for the hazard ratios of the non-terminal and terminal events, $\mbox{HR}_{NT}$ and $\mbox{HR}_{T}$, respectively. The number of replications in both simulation studies was 1000. The simulation algorithm is provided below.

\begin{enumerate}
\item We generate the binary covariates, $W_k$, where $k=1,...,p$ for $p$ covariates, in proportions that represent the renal transplant data.

\item We generate Pearson's correlation coefficient for the Normal copula, or association parameters for the Clayton, Frank and Gumbel copulas, respectively, as described in Table \ref{linksfortheta}, with chosen $b_k$.

\item Hazard rates for the non-terminal event, $\lambda_1$, and terminal event $\lambda_2$, with chosen $a_k$ and $c_k$ being generated by using equations \eqref{c5_l1_multi} and \eqref{c5_l2_multi}, respectively.

\item We use the conditional distribution method  \citep{Nelsen2006} to simulate time to the non-terminal and terminal events from a specified copula.

\begin{enumerate}
\item  $(U,V)$ have the joint distribution function of the chosen copula, where $U$ represents the uniformly transformed time to the non-terminal event, and $V$ represents the uniformly transformed time to the terminal event.
\item Generate $U$ from $\mbox{Uniform}(0,1)$.

\item Generate $V$ from $C(v|u)$, which is the inverse of the conditional copula distribution function of $V$ given $U$.
\end{enumerate}

\item We obtain $T_1$ and $T_2$ from the respective inverse marginal survival functions,
$$T_1=S_{T_1}^{-1}(U),$$
$$T_2=S_{T_2}^{-1}(V).$$

Here $\displaystyle T_1=\frac{-\log(U)}{\lambda_1}$ and $\displaystyle T_2=\frac{-\log(V)}{\lambda_2}$, when the survival distribution is Exponential for both events.

\item We simulate censoring time, $C$, independently from a Uniform distribution.

\item Set the time to the non-terminal event, $X=\min(T_1, T_2, C)$, with event indicator $d_1=1$ if $T_1 \leq \min(T_2,C)$ and $d_1=0$ otherwise.

\item Set the time to the terminal event, $Y=\min(T_2, C)$, with event indicator $d_2=1$ if $T_2 \leq C$ and $d_2=0$ otherwise.
\end{enumerate}

 In simulation study 1,  the simulated data are analysed using Cox model, and the models with underlying copula function. For example, when analysing the simulated data generated using the Clayton copula, we use the Clayton copula model. \cite{Sorrell2022} conducted an extensive simulation study to evaluate the effect of misspecification of copula functions and we will not duplicate that simulation in this paper.
\subsection{Choice of parameters}


We generate data sets with 3000 individuals with hazard rates and association parameters mimicking the real data analysis from the UKTR data set. The binary covariates are generated from a Bernoulli distribution with parameters mimicking the real data. Specifically, the age group  $>50$ years with probability 0.40, females with probability 0.38 and living donor recipients with probability 0.30.

\noindent The regression coefficients are set to be equal to those in Tables S1 and S2 in the Supplementary Materials for simulation studies 1 and 2, respectively. The censoring time, $C$, is generated independently from $\mbox{Uniform}(0,25)$ distribution, representing a maximum follow up time of 25 years.

\subsection{Performance measures}

We maximise the log-likelihood described in Section \ref{sec:meth} using the \texttt{optim} function in R to estimate the regression coefficients for the marginal and association parameters. The performance of the maximum likelihood estimates is evaluated using the bias, the mean squared error (MSE) and the coverage probability.


\subsection{Results of simulation study 1: performance of the proposed bivariate copula regression models}

\noindent For the non-terminal event, the hazard ratio of age group estimated from the Cox model has the following coverage probabilities,  $11.7\%$, $0.0\%$, $0.0\%$ and $18.9\%$, if data were simulated from Normal, Clayton, Frank and Gumbel copulas, respectively  (Table \ref{table:simulation1}). These coverage probabilities were far below the nominal level. In copula regression model with covariates included for the hazard rate but not for the association parameter (copula model 1), the equivalent coverage probabilities were $0.0\%$, $90.3\%$, $56.7\%$, $45.5\%$ (Table \ref{table:simulation1}). Using copula model 2, where covariates were included for the hazard rates and the association parameter, the coverage probability is around the nominal level for each copula, $94.6\%$, $95.1\%$, $94.5\%$, $95.2\%$ (Table \ref{table:simulation1}). We observe a reduction in the bias and MSE of the non-terminal hazard ratio for the age group in copula model 2, as compared to the Cox model. This explains the discrepancy in the findings between the Cox model and the copula model in the real data analysis.  \\

\noindent For the hazard ratios of sex and donor type for the non-terminal event, both the Cox model and the Copula model 2 result in coverage probabilities close to the nominal level. However, the results from the copula models show slight reductions in bias and MSE. For example, when using the Cox model for data generated from the Normal copula, the bias and MSE of the hazard ratio are $0.240$ and $0.062$, respectively, for the non-terminal event with covariate age agre. Comparing this to Copula model 2, we find the bias to be $-0.003$ and the MSE to be $0.007$. \\

\noindent The results for the hazard ratios of the terminal event are similar between the Cox model, Copula models 1 and 2 with coverage probability close to $95\%$. This is expected, because the terminal event can not be censored by the non-terminal event. However, using copula model 1 the coverage probability is $100.0$\%, for the hazard ratio of age group, where data are generated from the Clayton copula. When including the covariate age group in the association parameter, in copula model 2, we find the coverage probability increases to $95.7$\%.

\begin{sidewaystable*}[ht!]
\renewcommand{\arraystretch}{0.97} 
\centering
\small

\begin{tabular}{ |l  r r r  r r r  r r r  r r r |}
\hline
  & \multicolumn{3}{c}{Normal} & \multicolumn{3}{c}{Clayton} & \multicolumn{3}{c}{Frank} & \multicolumn{3}{c|}{Gumbel} \\
\hline
  Estimand                                    & Bias   & CP    &    MSE   & Bias   & CP    &    MSE   & Bias   & CP     &    MSE   & Bias   & CP & MSE\\
\hline
\multicolumn{13}{|l|}{\textbf{Cox Model}} \\
$\mbox{HR}_{NT},\mbox{age}$       &	0.240   &	11.7&	0.062&	0.552   &	0.0 &	0.309&	0.505   &	0.0&	0.260&	0.223   &	18.9&	0.055\\
$\mbox{HR}_{NT},\mbox{sex}$       &	$-$0.016&	94.3&	0.006&	$-$0.012&	95.2&	0.006&	$-$0.026&	94.0&	0.007&	$-$0.013&	94.8&	0.006\\
$\mbox{HR}_{NT},\mbox{donor}$     &	$-$0.019&	92.8&	0.003&	$-$0.019&	94.0&	0.003&	$-$0.024&	93.0&	0.003&	$-$0.006&	94.4&	0.003\\
$\mbox{HR}_T,\mbox{age}$        &	$-$0.010&	95.8&	0.079&	0.000   &	94.6&	0.086&	$-$0.012&	95.4&	0.080&	$-$0.004&	96.4&	0.074\\
$\mbox{HR}_T,\mbox{sex}$        &	$-$0.001&	95.1&	0.005&	$-$0.005&	94.7&	0.005&	$-$0.005&	94.8&	0.005&	0.000   &	95.5&	0.005\\
$\mbox{HR}_T,\mbox{donor}$      &	0.002   &	95.2&	0.002&	0.000   &	94.9&	0.002&	$-$0.001&	95.5&	0.002&	0.001   &	94.8&	0.002\\

\hline
\multicolumn{13}{|l|}{\textbf{Copula model 1}} \\


$\mbox{HR}_{NT},\mbox{age}$     &	$-$0.155  &	0.000&	0.024&	$-$0.042&	90.3&	0.012&	$-$0.147 &	56.7&	0.030&	$-$0.146 &	45.5&	0.027\\
$\mbox{HR}_{NT},\mbox{sex}$     &	0.075     &	100.0&	0.006&	0.004   &	95.0&	0.005&	0.007    &	94.9&	0.005&	0.004    &	94.6&	0.005\\
$\mbox{HR}_{NT},\mbox{donor}$   &	$-$0.018  &	100.0&	0.000&	$-$0.007&	92.5&	0.003&	$-$0.002 &	94.0&	0.002&	0.000    &	94.2&	0.003\\
$\mbox{HR}_T,\mbox{age}$        &	$-$0.042  &	100.0&	0.002&	0.397   &	76.9&	0.252&	0.144    &	94.6&	0.105&	$-$0.132 &	91.2&	0.083\\
$\mbox{HR}_T,\mbox{sex}$        &	0.046     &	100.0&	0.002&	0.017   &	95.4&	0.004&	$-$0.001 &	95.3&	0.004&	$-$0.002 &	94.9&	0.004\\
$\mbox{HR}_T,\mbox{donor}$      &	$-$0.051  &	100.0&	0.003&	0.028   &	92.4&	0.003&	0.003    &	95.6&	0.002&	$-$0.003 &	94.6&	0.002\\

\hline
\multicolumn{13}{|l|}{\textbf{Copula model 2}}\\
$\mbox{HR}_{NT},\mbox{age}$     &	$-$0.003  &	94.6&	0.007&	$-$0.002 &	95.1&	0.009&	$-$0.002 &	94.5&	0.010&	$-$0.003 &	95.2&	0.007\\
$\mbox{HR}_{NT},\mbox{sex}$     &	$-$0.003  &	94.6&	0.006&	$-$0.001 &	95.9&	0.005&	$-$0.002 &	95.3&	0.005&	$-$0.003&	95.4&	0.006\\
$\mbox{HR}_{NT},\mbox{donor}$   &	0.002     &	94.1&	0.003&	0.002    &	93.6&	0.002&	0.002    &	94.0&	0.003&	0.002    &	94.4&	0.003\\
$\mbox{HR}_T,\mbox{age}$        &	$-$0.009  &	95.7&	0.077&	$-$0.011 &	95.1&	0.080&	$-$0.014 &	95.0&	0.093&	$-$0.004 &	95.9&	0.073\\
$\mbox{HR}_T,\mbox{sex}$        &	$-$0.001  &	95.1&	0.004&	$-$0.002 &	95.5&	0.004&	$-$0.001 &	94.6&	0.005&	0.000    &	94.8&	0.004\\
$\mbox{HR}_T,\mbox{donor}$      &	0.002     &	95.1&	0.002&	0.001    &	94.0&	0.002&	0.002    &	94.6&	0.002&	0.002    &	94.7&	0.002\\
$\beta_0$                       &	0.005     &	94.3&	0.005&	0.017    &	95.5&	0.031&	0.018    &	94.2&	0.253&	0.066    &	96.6&	0.123\\
$\beta_1$                       &	$-$0.006  &	96.0&	0.006&	$-$0.018 &	96.2&	0.032&	$-$0.056 &	95.6&	0.557&	$-$0.055 &	96.9&	0.129\\
$\beta_2$                       &	$-$0.004  &	95.4&	0.007&	$-$0.004 &	95.1&	0.025&	$-$0.033 &	94.9&	0.445&	$-$0.007 &	96.4&	0.077\\
$\beta_3$                       &	0.002     &	96.0&	0.009&	$-$0.005 &	94.6&	0.033&	$-$0.030 &	95.0&	0.839&	0.014    &	95.6&	0.094\\

\hline
\end{tabular}
\caption{Results of simulation study. Copula model 1 represents the copula model with  the marginal hazard rates depending on the covariates. Copula model 2 represents the copula model with both the marginal hazard rates and association parameters depending on the covariates. Data are generated from the Normal, Clayton, Frank and Gumbel copulas and the true copula distribution is used for estimation. We generate 1000 data sets with 3000 individuals in each. The non-terminal hazard ratio for each covariate is given by $\mbox{HR}_{NT,\mbox{covariate}}$, and the terminal hazard ratio for each covariate is given by $\mbox{HR}_{T, \mbox{covariate}}$. MSE refers to the mean squared error and CP refers to the coverage probability, given as a percentage.}
\label{table:simulation1}
\end{sidewaystable*}

\subsection{Results of simulation study 2: effects of the misspecification of survival distributions}

We evaluate the use of alternative survival distributions and the effects of misspecification of the survival distribution.
For the simulation studies investigating misidentification, we simulate data mimicking the real data set using four different copula distributions, with true values given in supplementary material Table S2. We also evaluate the use of AIC to select the survival distributions. We simulate data with the following combinations of survival distributions and copula functions,  Exponential, Weibull and Gompertz survival distributions and Normal, Clayton, Gumbel and Frank copulas.

The results of simulation study 2 are given in Table S3 in the Supplementary Material. Here we give a summary of finding from this simulation study. Our simulation study  shows that AIC can be used to select the survival distributions. The survival distribution is chosen correctly in almost 100\% cases for the Weibull distribution, roughly 91\% or above for Gompertz distribution, and around 78\% or above for Exponential distribution. Where the underlying Exponential distribution is not correctly identified, about 10\%  of the times the survival distribution is misspecified as Weibull or Gompertz distribution. However, the estimation is in general robust to the misidentification of the Exponential distributions by Weibull or Gompertz distribution. The summary of performance of the models selected by lowest AIC show that the coverage probability is close to the nominal level.

\section{Discussion}\label{sec:conc}

We propose a set of copula regression models to allow both the hazard rates and the association between times to non-terminal and terminal events to vary by covariates. We use conditional copulas \citep{Patton2006b} with predictors for the hazard rates of the semi-competing risk events and the association parameters.  The advantage of the proposed method is the estimation of the hazard ratios by taking into account the correlation between the semi-competing risks and the flexibility of using a variety of copulas to describe different patterns in the relationship between the survival endpoints. \\

\noindent Our work demonstrates the importance of considering the correlation between the semi-competing risks.  In the real data analysis, different conclusions were found for the effect of age group, where Cox model showed older age was associated with lower risk of graft failure (HR: 0.911, 95\%CI: 0.872 to 0.952), while our proposed copula models showed older age was associated with higher risk of graft failure from all copula models. The estimated hazard ratio from Frank copula Weibull survival model, which has the lowest AIC, is 1.202 with 95\%CI: 1.152 to 1.251 (Table \ref{table:hr_allcovs_Weibull}).

\noindent We conduct simulation studies to assess the performance of the copula regression models and to evaluate the use of AIC to choose survival distributions. The estimation of the hazard ratio for the non-terminal event is improved by using the copula model 2, with coverage probability around the nominal level, compared to using the Cox model which has coverage probability 0\% in some scenarios (Table \ref{table:simulation1}). These results corroborate the findings from \cite{Leffondre2013}. For the non-terminal hazard ratio, we found a reduction in bias and MSE using the copula regression models compared to the Cox model. Our work highlights the importance of acknowledging the semi-competing risk when analysing the effect of a covariate on an endpoint, where the covariate has a strong effect on a competing risk.  \\

\noindent We have considered the effect of misspecification of survival distribution. Our simulation studies show that the AIC can be used to select the survival distributions in the majority of cases. In the remaining cases, misspecification is more likely to occur when the underlying distribution is Exponential or Gompertz. However, miss-specifying these two distributions by a Weibull distribution still holds good properties in terms of the parameter estimates. Further research may also investigate other parametric survival distributions or the use of non-parametric methods to model the marginal survival functions.

\noindent We have used a full maximum likelihood approach for estimating the model parameters. We acknowledge the two-stage estimation procedures for the association parameter may be more efficient for copula survival models \citep{Shih1995}. In the two-stage approach, the association parameter and the parameters in the marginal survival distributions are estimated separately. In stage 1, parameters in the marginal distributions are estimated assuming they are independent. In stage 2, the association parameter is estimated by fixing the two marginal distributions at the estimates from stage 1. This approach ignores the dependency structure in stage 1, however, it offers the advantage of being practically efficient especially when the models become more complex. The application of the two-stage approach in our proposed models and the comparison with the one-stage full maximum likelihood approach is a topic for future research. Another possible extension could be to estimate the marginal survival function using non-parametric methods \citep{Li2020} or Cox proportional hazards model \citep{Li2021}. It would be also of interest to investigate the use of grid search methods to find starting values for optimising the likelihood function.

\noindent In this paper, we have extended the recent work in bivariate semi-competing risk models by including covariates. Another topic of future research could be to extend the model to include frailty terms to account for unmeasured covariates. Since we have bivariate semi-competing risk data, this extension will require to model the frailty terms by using a bivariate distribution to account for the potential correlation between the two frailty terms.

\noindent We have used Exponential, Weibull and Gompertz distributions to illustrate the methods and applications of our proposed models. However, this can be readily extended to other parametric survival distributions. We have assessed the performance of our proposed copula survival models using simulation studies and compare between them using AIC. Assessing the goodness-of-fit of the copula survival models may be developed in future research.

\noindent As this paper focuses on describing copula regression models with binary covariates, the dichotomous age groups ($>50$ years or $\le 50$ years) representing younger and older age were included for illustration purpose. We acknowledge that age should be best analysed with more categories or as a continuous variable. In our follow-up work, we are developing copula regression models to include more categories for age or include age as a continuous variable.

\noindent Further research may investigate the inclusion of continuous and categorical covariates. We have considered a linear function to incorporate covariates into the hazard rates and association parameter, however, alternatives may be considered. \cite{Acar2013} have developed methods to compare potential functions that describe the association parameter of the copula function's relationship with the covariate by using a generalised likelihood ratio test. We have used available case analysis to illustrate the application of our methods. Future research can use multiple imputation to deal with missing data when using our proposed methods. This could be time consuming when the Normal copula model is used.

\noindent As can be seen in Tables \ref{table:hr_allcovs}-\ref{table:hr_allcovs_Weibull}, it takes 2-4 hours to optimize the likelihood for Normal copula models. Future research may investigate how to speed up the optimization process for the analysis of a single data set, and the use of parallel computing in R for conducting simulation studies.

\subsubsection*{Acknowledgments}
The authors gratefully acknowledge the NHS Blood and Transplant for the access to the UK Transplant Registry data used in this project. We are grateful to all the transplant centres in the UK who contributed data on which this project is based. LS received an University of Plymouth School of Engineering, Computing and Mathematics PhD studentship. YW was supported by an UKRI MRC Fellowship (MC/W021358/1) and received funding from UKRI EPSRC Impact Acceleration Account (EP/X525789/1). We are grateful to have access to the University of Plymouth Faculty of Science and Engineering High Performance Computing Clusters for conducting our simulation studies.

\bibliographystyle{apa}
\bibliography{PhD}

\end{document}


\begin{center}
Supplementary Material for \textbf{``Bivariate copula regression models for semi-competing risks"}

by Yinghui Wei, Ma{\l}gorzata Wojty{\'s}, Lexy Sorrell and Peter Rowe

\end{center}

%
%
%
%


\section{The Delta method}
\label{appendix:delta}

\noindent Based on the Delta method (\cite{vandervaart2000}, p. 25), for a univariate function of a random variable $X$, $f(X)$, the variance can be approximated by
\begin{equation}\label{var_approx1}
\mbox{Var}[f(X)] =\left(f'(\mu_x)\right)^2 \mbox{Var}(X),
\end{equation}
where $\mu_x = E(X)$.

\noindent For a bivariate function of random variables $X$ and $Y$, $f(X,Y)$, the variance can be approximated by
\begin{equation}\label{var_approx2}
\mbox{Var}[f(X,Y)] = \left(f'_x(\bm{\mu})\right)^2 \mbox{Var}(X)  + 2f'_x(\bm{\mu})f'_y(\bm{\mu})\mbox{Cov}(X,Y)+ \left(f'_y(\bm{\mu})\right)^2\mbox{Var}(Y),
\end{equation}
where $\bm{\mu} = (\mu_x, \mu_y)=(E(X), E(Y))$.

\noindent For a $p$-variate function $f(X_1, \ldots, X_p)$, the variance can be approximated by
\begin{equation}\label{var_approx3}
\mbox{Var}[f(X_1, \ldots, X_p)] =  \nabla f(\bm{\mu})^{T} \boldsymbol\Sigma_{\bm X} \nabla f(\bm{\mu}),
\end{equation}
where $\boldsymbol\Sigma_{\bm X}$ is the covariance matrix of the random vector ${\bm X}=(X_1, \ldots, X_p)$ and $\nabla f(\bm{\mu})$ is the gradient of $f$ with $\bm{\mu} = (E(X_1), \ldots, E(X_p))$.

\subsection{Variance of the hazard ratio}\label{var_hr}

To calculate the variance of hazard ratios, we use the Delta method. For the Exponential, Weibull and Gompertz distributions, the hazard ratio is given in the same exponential form. Thus, we find the approximation of the variance of $f(X)=\exp(X)$, which is given by

$$\mbox{Var}[f(X)] = \exp(2\mu) \mbox{Var}(X).$$

\subsection{Variance of the Normal association parameter}\label{var_ap_normal}

\subsubsection{One covariate}

When the Normal copula parameter ${\rho}$ varies with covariate, $W_1$, we assume \\$\displaystyle {\rho}=\frac{\exp(2(b_0+b_1 W_1))-1}{\exp(2(b_0+b_1 W_1))+1}$ therefore we want an approximation for the variance of the function $\displaystyle f(b_0,b_1)=\frac{\exp(2(b_0+b_1 W_1))-1}{\exp(2(b_0+b_1 W_1))+1}$. The derivatives are given by the following,

\begin{equation}\label{app_normal_1}
f'_{\beta_0} = \frac{\partial}{\partial b_0}\left(\frac{\exp(2(b_0+b_1 W_1))-1}{\exp(2(b_0+b_1 W_1))+1}\right)=\frac{4\exp(2(b_0+b_1 W_1))}{(\exp(2(b_0+b_1W_1))+1)^2},
\end{equation}

\begin{equation}\label{app_normal_2}
f'_{b_1} = \frac{\partial}{\partial b_1}\left(\frac{\exp(2(b_0+b_1 W_1))-1}{\exp(2(b_0+b_1 W_1))+1}\right)= \frac{4W_1\exp(2(b_0+b_1 W_1))}{(\exp(2(b_0+b_1W_1))+1)^2}.
\end{equation}

\noindent Substituting \eqref{app_normal_1} and \eqref{app_normal_2} into equation \eqref{var_approx2}, the variance is approximated by,


\begin{equation}
\begin{split}
\mbox{Var}(\hat{\rho})&= \frac{16 \exp(4({b}_0+{b}_1 W_1))}{(\exp(2({b}_0+{b}_1 W_1))+1)^4} \left[\mbox{Var}(\hat{b}_0) + 2 W_1 \mbox{Cov}(\hat{b}_0,\hat{b}_1)+W_1^2\mbox{Var}(\hat{b}_1) \right].
\end{split}
\end{equation}

\subsubsection{Multiple covariates}

\noindent In general, when $p$ covariates $W_1,\ldots, W_p$ are considered, we assume \\$\displaystyle {\rho}=\frac{\exp(2({b}_0+{b}_1 W_1 + \ldots + {b}_p W_p))-1}{\exp(2({b}_0+{b}_1 W_1 \ldots + {b}_p W_p))+1}$. Using vector notation, we define the function
$$
\displaystyle f(\boldsymbol{b})=\frac{\exp\left(2\boldsymbol{b}^T {\bm W}\right)-1}{\exp\left(2\boldsymbol{b}^T {\bm W}\right)+1},
$$
where $\boldsymbol{b} = ({b}_0,{b}_1, \ldots, {b}_p)^T$ and ${\bm W}=(1, W_1,\ldots, W_p)^T$.

\noindent The partial derivative with respect to ${b}_k$ is given by
\begin{equation}\label{der1}
 f'_{{b}_k} = \frac{\partial}{\partial {b}_k} f(\boldsymbol{b} ) = \frac{4 W_k \exp(2 \boldsymbol{b}^T {\bm W})}{(\exp(2 \boldsymbol{b}^T {\bm W}) + 1)^2}
\end{equation}
for $k=0,1,\ldots, p$. Therefore, using equation (\ref{var_approx2}), the variance of $\hat{\rho}$ is approximated by
$$
  \mbox{Var}(\hat\theta) =
  \nabla f({\boldsymbol{b}})^T \boldsymbol\Sigma_{\hat{\boldsymbol{b}}} \nabla f({\boldsymbol{b}}),
$$
where $\boldsymbol\Sigma_{\hat{\boldsymbol{b}}}$ is the covariance matrix of $\hat{\boldsymbol{b}}$ and
$\nabla f({\boldsymbol{b}}) = \left(f'_{{b}_0}, f'_{{b}_2}, \ldots, f'_{{b}_p}  \right)^T$ as given in (\ref{der1}).

\subsection{Variance of the Clayton association parameter}\label{var_ap_clayton}

\subsubsection{One covariate}

For the Clayton copula, we assume $\theta=\exp({b}_0+{b}_1  W_1)$, therefore we want an approximation for the variance of the function $f({b}_0,{b}_1)=\exp({b}_0+{b}_1 W_1)$. The derivatives are given by,

\begin{equation}\label{app_clayton_1}
f'_{{b}_0} = \frac{\partial \exp({b}_0+{b}_1 W_1)}{\partial {b}_0}=\exp({b}_0+{b}_1 W_1),
\end{equation}

\begin{equation}\label{app_clayton_2}
f'_{{b}_1} = \frac{\partial \exp({b}_0+{b}_1 W_1)}{\partial {b}_1}=W_1 \exp({b}_0+{b}_1 W_1).
\end{equation}


\noindent Substituting \eqref{app_clayton_1} and \eqref{app_clayton_2} into equation \eqref{var_approx2}, the variance is approximated by,


\begin{equation}
\begin{split}
\mbox{Var}(\hat\theta)&=
\exp(2({b}_0+{b}_1 W_1)) [\mbox{Var}(\hat{b}_0) + 2W_1\mbox{Cov}(\hat{b}_0,\hat{b}_1)+W_1^2\mbox{Var}(\hat{b}_1)].
\end{split}
\end{equation}

\subsubsection{Multiple covariates}

\noindent In general, when $p$ covariates $W_1,\ldots, W_p$ are considered, we assume \\$\displaystyle \theta=\exp({b}_0+{b}_1 W_1 + \ldots + {b}_p W_p)$. Using vector notation, we define the function
$$
\displaystyle f(\boldsymbol{b})=\exp\left(\boldsymbol{b}^T {\bm W}\right),
$$
where $\boldsymbol{b} = ({b}_0,{b}_1, \ldots, {b}_p)^T$ and ${\bm W}=(1, W_1,\ldots, W_p)^T$.

\noindent The partial derivative with respect to ${b}_k$ is given by
\begin{equation}\label{der2}
 f'_{{b}_k} = \frac{\partial}{\partial {b}_k} f(\boldsymbol{b} ) = W_k \exp( \boldsymbol{b}^T {\bm W})
\end{equation}
for $k=0,1,\ldots, p$. Therefore, using equation (\ref{var_approx2}), the variance of $\hat\theta$ is approximated by
$$
  \mbox{Var}(\hat{\rho}) =
  \nabla f({\boldsymbol{b}})^T \boldsymbol\Sigma_{\hat{\boldsymbol{b}}} \nabla f({\boldsymbol{b}}),
$$
where $\boldsymbol\Sigma_{\hat{\boldsymbol{b}}}$ is the covariance matrix of $\hat{\boldsymbol{b}}$ and
$\nabla f({\boldsymbol{b}}) = \left(f'_{{b}_0}, f'_{{b}_2}, \ldots, f'_{{b}_p}  \right)^T$ as given in (\ref{der2}).

\subsection{Variance of the Gumbel association parameter}\label{var_ap_gumbel}

\subsubsection{One covariate}

For the Gumbel copula, we assume $\theta=\exp({b}_0+{b}_1  W_1)+1$ therefore we want an approximation for the variance of the function \\ $f({b}_0,{b}_1)=\exp({b}_0+{b}_1 W_1)+1$. The derivatives are given by,

\begin{equation}\label{app_gumbel_1}
f'_{{b}_0} = \frac{\partial \exp({b}_0+{b}_1 W_1)+1}{\partial {b}_0}=\exp({b}_0+{b}_1 W_1),
\end{equation}

\begin{equation}\label{app_gumbel_2}
f'_{{b}_1} = \frac{\partial \exp({b}_0+{b}_1 W_1)+1}{\partial {b}_1}=W_1 \exp({b}_0+{b}_1 W_1).
\end{equation}

\noindent Substituting \eqref{app_gumbel_1} and \eqref{app_gumbel_2} into equation \eqref{var_approx2}, the variance is approximated by,


\begin{equation}
\begin{split}
\mbox{Var}(\hat\theta)&=
\exp(2({b}_0+{b}_1 W_1)) [\mbox{Var}(\hat{b}_0) + 2W_1\mbox{Cov}(\hat{b}_0,\hat{b}_1)+W_1^2\mbox{Var}(\hat{b}_1)].
\end{split}
\end{equation}

\subsubsection{Multiple covariates}

\noindent In general, when $p$ covariates $W_1,\ldots, W_p$ are considered, we assume \\$\displaystyle \theta=\exp({b}_0+{b}_1 W_1 + \ldots + {b}_p W_p)+1$. Using vector notation, we define the function
$$
\displaystyle f(\boldsymbol{b})=\exp\left(\boldsymbol{b}^T {\bm W}\right)+1,
$$
where $\boldsymbol{b} = ({b}_0,{b}_1, \ldots, {b}_p)^T$ and ${\bm W}=(1, W_1,\ldots, W_p)^T$.

\noindent The partial derivative with respect to ${b}_k$ is given by
\begin{equation}\label{der3}
 f'_{{b}_k} = \frac{\partial}{\partial {b}_k} f(\boldsymbol{b} ) = W_k \exp( \boldsymbol{b}^T {\bm W})
\end{equation}
for $k=0,1,\ldots, p$. Therefore, using equation (\ref{var_approx2}), the variance of $\hat\theta$ is approximated by
$$
  \mbox{Var}(\hat{\rho}) =
  \nabla f({\boldsymbol{b}})^T \boldsymbol\Sigma_{\hat{\boldsymbol{b}}} \nabla f({\boldsymbol{b}}),
$$
where $\boldsymbol\Sigma_{\hat{\boldsymbol{b}}}$ is the covariance matrix of $\hat{\boldsymbol{b}}$ and
$\nabla f({\boldsymbol{b}}) = \left(f'_{{b}_0}, f'_{{b}_2}, \ldots, f'_{{b}_p}  \right)^T$ as given in (\ref{der3}).

\begin{landscape}
\section{Regression coefficients estimated from copula regression models}

\begin{table}[ht!]
\centering
\begin{tabular}{|crrrr|}
\hline
Parameter          & Normal-Exponential & Clayton-Exponential & Frank-Exponential& Gumbel-Exponential \\
\hline
\multicolumn{5}{|l|}{Graft failure} \\
$a_0$	&$-$3.30 ($-$3.33, $-$3.27)	&$-$3.28 ($-$3.31, $-$3.24)	&$-$3.28 ($-$3.31, $-$3.24)	&$-$3.33 ($-$3.36, $-$3.29)\\
$a_1$	& 0.11 ( 0.07,  0.16)	& 0.32 ( 0.29,  0.36)	& 0.31 ( 0.27,  0.35)	& 0.13 ( 0.08,  0.17)\\
$a_2$	& 0.01 ($-$0.03,  0.06)	& 0.01 ($-$0.03,  0.05)	& 0.00 ($-$0.04,  0.04)	& 0.01 ($-$0.03,  0.06)\\
$a_3$	&$-$0.51 ($-$0.56, $-$0.46)	&$-$0.53 ($-$0.58, $-$0.48)	&$-$0.53 ($-$0.58, $-$0.48)	&$-$0.51 ($-$0.56, $-$0.46)\\
\hline
\multicolumn{5}{|l|}{Death} \\
$c_0$	&$-$4.15 ($-$4.19, $-$4.11)	&$-$4.09 ($-$4.13, $-$4.05)	&$-$4.08 ($-$4.12, $-$4.04)	&$-$4.16 ($-$4.20, $-$4.11)\\
$c_1$	& 1.32 ( 1.27,  1.37)	& 1.35 ( 1.31,  1.40)	& 1.35 ( 1.31,  1.40)	& 1.30 ( 1.25,  1.35)\\
$c_2$	&$-$0.11 ($-$0.15, $-$0.06)	&$-$0.07 ($-$0.11, $-$0.03)	&$-$0.07 ($-$0.11, $-$0.03)	&$-$0.11 ($-$0.15, $-$0.06)\\
$c_3$	&$-$0.65 ($-$0.71, $-$0.59)	&$-$0.62 ($-$0.67, $-$0.56)	&$-$0.63 ($-$0.68, $-$0.57)	&$-$0.64 ($-$0.70, $-$0.58)\\
\hline
\multicolumn{5}{|l|}{Association} \\
$b_0$	& 0.35 ( 0.32,  0.38)	& 0.39 ( 0.29,  0.50)	& 3.04 ( 2.75,  3.33)	&$-$2.30 ($-$2.44, $-$2.15)\\
$b_1$	& 0.28 ( 0.25,  0.32)	& 1.09 ( 0.98,  1.20)	& 5.07 ( 4.60,  5.54)	& 1.35 ( 1.20,  1.51)\\
$b_2$	& 0.02 ($-$0.01,  0.06)	& 0.14 ( 0.04,  0.24)	& 0.35 ($-$0.06,  0.76)	& 0.06 ($-$0.08,  0.19)\\
$b_3$	& 0.03 ($-$0.03,  0.08)	& 0.53 ( 0.38,  0.67)	& 0.86 ( 0.21,  1.51)	&$-$0.04 ($-$0.23,  0.15)\\
\hline
\end{tabular}
\caption{Regression coefficients estimated from copula exponential survival models. Here $a_i (i=0, 1, 2, 3)$ are regression coefficients for the hazard function for graft failure,
$c_i (i=0,1,2,3)$ are regression coefficients for the hazard function for death, and $b_i (i=0,1,2,3)$ are regression coefficients for the association parameter between graft failure and death.}
\label{table:est_coef_real_exp}
\end{table}

\begin{table}[ht!]
\centering
\begin{tabular}{|crrrr|}
\hline
Parameter          & Normal-Gompertz & Clayton-Gompertz & Frank-Gompertz& Gumbel-Gompertz\\
\hline
\multicolumn{5}{|l|}{Graft failure} \\
$a_0$	& 0.02 ($-$0.02,  0.06)	& 0.00 ($-$0.04,  0.04)	& 0.00 ($-$0.04,  0.04)	& 0.01 ($-$0.03,  0.05)\\
$a_1$	&$-$0.52 ($-$0.57, $-$0.47)	&$-$0.52 ($-$0.57, $-$0.47)	&$-$0.61 ($-$0.66, $-$0.56)	&$-$0.52 ($-$0.57, $-$0.47)\\
$a_2$	& 0.34 ( 0.31,  0.37)	& 0.41 ( 0.31,  0.51)	& 3.92 ( 3.62,  4.22)	&$-$2.31 ($-$2.45, $-$2.17)\\
$a_3$	& 0.28 ( 0.25,  0.32)	& 0.95 ( 0.84,  1.06)	& 2.07 ( 1.66,  2.49)	& 1.40 ( 1.25,  1.55)\\
\hline
\multicolumn{5}{|l|}{Death} \\
$c_0$	& 0.02 ($-$0.01,  0.06)	& 0.16 ( 0.06,  0.26)	& 0.64 ( 0.23,  1.04)	&$-$0.02 ($-$0.15,  0.12)\\
$c_1$	& 0.02 ($-$0.03,  0.07)	& 0.29 ( 0.12,  0.46)	& 0.42 ($-$0.24,  1.09)	&$-$0.10 ($-$0.29,  0.10)\\
$c_2$	&$-$4.56 ($-$4.61, $-$4.50)	&$-$4.33 ($-$4.38, $-$4.28)	&$-$4.33 ($-$4.38, $-$4.28)	&$-$4.57 ($-$4.63, $-$4.52)\\
$c_3$	& 1.43 ( 1.38,  1.47)	& 1.37 ( 1.32,  1.42)	& 1.37 ( 1.33,  1.42)	& 1.41 ( 1.37,  1.46)\\
\hline
\multicolumn{5}{|l|}{Association} \\
$b_0$	&$-$0.11 ($-$0.16, $-$0.07)	&$-$0.10 ($-$0.14, $-$0.06)	&$-$0.08 ($-$0.12, $-$0.04)	&$-$0.12 ($-$0.16, $-$0.07)\\
$b_1$	&$-$0.59 ($-$0.65, $-$0.53)	&$-$0.59 ($-$0.64, $-$0.53)	&$-$0.70 ($-$0.76, $-$0.64)	&$-$0.58 ($-$0.64, $-$0.52)\\
$b_2$	&$-$0.02 ($-$0.02, $-$0.01)	& 0.00 ( 0.00,  0.00)	&$-$0.01 ($-$0.01, $-$0.01)	&$-$0.02 ($-$0.03, $-$0.02)\\
$b_3$	& 0.05 ( 0.05,  0.06)	& 0.03 ( 0.03,  0.04)	& 0.04 ( 0.04,  0.04)	& 0.06 ( 0.05,  0.06)\\
\hline
\multicolumn{5}{|l|}{Shape parameter} \\
$\gamma_1$	&$-$3.20 ($-$3.24, $-$3.16)	&$-$3.26 ($-$3.30, $-$3.21)	&$-$3.19 ($-$3.23, $-$3.14)	&$-$3.20 ($-$3.25, $-$3.16)\\
$\gamma_2$	& 0.08 ( 0.03,  0.12)	& 0.26 ( 0.22,  0.30)	& 0.21 ( 0.17,  0.25)	& 0.06 ( 0.02,  0.11)\\
\hline
\end{tabular}
\caption{Regression coefficients estimated from copula Gompertz survival models. Here $a_i (i=0, 1, 2, 3)$ are regression coefficients for the hazard function for graft failure,
$c_i (i=0,1,2,3)$ are regression coefficients for the hazard function for death, $b_i (i=0,1,2,3)$ are regression coefficients for the association parameter between graft failure and death,  and $\gamma_j (j=1,2)$ are the shape parameters in Gompertz distribution.}
\label{table:est_coef_real_gompertz}
\end{table}

\begin{table}[ht!]
\centering
\begin{tabular}{|c  rrrr|}
\hline
Parameter          & Normal-Weibull & Clayton-Weibull & Frank-Weibull& Gumbel-Weibull \\
\hline
 \multicolumn{5}{|l|}{Graft failure} \\
$a_0$	&$-$2.51 ($-$2.56, $-$2.47)	&$-$2.58 ($-$2.62, $-$2.54)	&$-$2.56 ($-$2.60, $-$2.52)	&$-$2.56 ($-$2.60, $-$2.51)\\
$a_1$	& 0.01 ($-$0.04,  0.05)	& 0.19 ( 0.15,  0.23)	& 0.18 ( 0.14,  0.22)	&$-$0.01 ($-$0.05,  0.03)\\
$a_2$	& 0.02 ($-$0.02,  0.06)	& 0.02 ($-$0.02,  0.07)	& 0.01 ($-$0.03,  0.05)	& 0.02 ($-$0.02,  0.06)\\
$a_3$	&$-$0.56 ($-$0.61, $-$0.51)	&$-$0.58 ($-$0.63, $-$0.53)	&$-$0.58 ($-$0.63, $-$0.53)	&$-$0.54 ($-$0.59, $-$0.49)\\

\hline
\multicolumn{5}{|l|}{Death} \\
$c_0$	&$-$4.14 ($-$4.20, $-$4.08)	&$-$4.03 ($-$4.09, $-$3.96)	&$-$4.03 ($-$4.09, $-$3.97)	&$-$4.00 ($-$4.06, $-$3.94)\\
$c_1$	& 1.31 ( 1.26,  1.35)	& 1.31 ( 1.26,  1.35)	& 1.32 ( 1.27,  1.36)	& 1.28 ( 1.24,  1.33)\\
$c_2$	&$-$0.10 ($-$0.14, $-$0.06)	&$-$0.08 ($-$0.12, $-$0.04)	&$-$0.08 ($-$0.12, $-$0.04)	&$-$0.10 ($-$0.14, $-$0.06)\\
$c_3$	&$-$0.65 ($-$0.71, $-$0.59)	&$-$0.64 ($-$0.70, $-$0.59)	&$-$0.65 ($-$0.71, $-$0.60)	&$-$0.65 ($-$0.71, $-$0.59)\\
\hline
\multicolumn{5}{|l|}{Association} \\
$b_0$	& 0.42 ( 0.38,  0.46)	& 0.41 ( 0.29,  0.53)	& 3.20 ( 2.89,  3.52)	&$-$1.81 ($-$1.94, $-$1.67)\\
$b_1$	& 0.28 ( 0.24,  0.33)	& 0.91 ( 0.78,  1.03)	& 3.95 ( 3.49,  4.42)	& 1.07 ( 0.94,  1.21)\\
$b_2$	& 0.03 ($-$0.01,  0.07)	& 0.16 ( 0.05,  0.27)	& 0.17 ($-$0.25,  0.59)	& 0.04 ($-$0.08,  0.17)\\
$b_3$	& 0.03 ($-$0.03,  0.09)	& 0.48 ( 0.31,  0.64)	& 0.49 ($-$0.17,  1.16)	&$-$0.04 ($-$0.22,  0.13)\\
\hline
\multicolumn{5}{|l|}{Shape parameter} \\
$\alpha_1$	& 0.67 ( 0.65,  0.68)	& 0.70 ( 0.69,  0.71)	& 0.70 ( 0.68,  0.71)	& 0.67 ( 0.66,  0.68)\\
$\alpha_2$	& 1.02 ( 1.00,  1.04)	& 0.98 ( 0.96,  0.99)	& 0.99 ( 0.97,  1.01)	& 0.96 ( 0.94,  0.98)\\
\hline
\end{tabular}
\caption{Regression coefficients estimated from copula Weibull survival models. Here $a_i (i=0, 1, 2, 3)$ are regression coefficients for the scale parameter in Weibull distribution for graft failure,
$b_i (i=0,1,2,3)$ are regression coefficients for the association parameter between graft failure and death, $c_i (i=0,1,2,3)$ are regression coefficients for the scale parameter in Weibull distribution for death and $\alpha_j (j=1,2)$ are the shape parameters in Weibull distribution.}
\label{table:est_coef_real_weibull}
\end{table}
\end{landscape}

\section{True values used in simulation studies}
\begin{table}[ht!]
\centering
\begin{tabular}{|crrrr|}
\hline
          & \multicolumn{4}{c|}{Copula}\\
Parameter          & Normal & Clayton & Frank & Gumbel\\
\hline
\multicolumn{5}{|l|}{Graft failure} \\
$a_0$ & $-$3.30   & $-$3.28  &  $-$3.37   & $-$3.33\\
$a_1$ &  0.11     & 0.32     &  0.31      & 0.13 \\
$a_2$ &   0.02    & 0.00     &  0.00      & 0.00\\
$a_3$ &  $-$0.51  & $-$0.53  &  $-$0.53   & $-$0.51 \\
\hline
\multicolumn{5}{|l|}{Death} \\
$c_0$ &   $-$4.15 &  $-$4.09 &  $-$4.08   & $-$4.16 \\
$c_1$ &   1.32    &  1.35    &  1.35      & 1.30 \\
$c_2$ &   $-$0.11 &  $-$0.07 &   $-$0.07  & $-$0.11 \\
$c_3$ &   $-$0.65 &  $-$0.62 &   $-$0.62  & $-$0.64 \\
\hline
\multicolumn{5}{|l|}{Association} \\
$b_0$ &   0.35    &  0.39    &   3.06     & $-$2.30 \\
$b_1$ &   0.28    &   1.09   &  5.07      & 1.35  \\
$b_2$ &   0.00    &   0.14   &  0.00      & 0.00\\
$b_3$ &   0.00    &   0.53   &  0.86      & 0.00 \\
\hline
\end{tabular}
\caption{True values used in simulation study 1: comparison of the Cox model and Copula models.}
\label{table:allcovs_real}
\end{table}

\begin{table}[ht!]
\centering
\begin{tabular}{|c  rrrr|}
\hline
            & \multicolumn{4}{c|}{Copula}\\
 Parameter           & Normal & Clayton & Frank & Gumbel\\
\hline
\multicolumn{5}{|l|}{Exponential survival distribution} \\
$a_0$ &  $-$3.440  & $-$3.420  & $-$3.420   & $-$2.240 \\
$a_1$ &  0.170  & 0.380     & 0.370      & 1.350 \\
$c_0$ &  $-$4.360  & $-$4.28  & $-$4.270   & $-$4.360 \\
$c_1$ &   1.390 &  1.410    & 1.410      & 1.360 \\
$b_0$ &  0.370  &  0.620    & 3.440      & $-$2.240 \\
$b_1$ &   0.290 &  1.040    & 5.230      &  1.350 \\
\hline
 \multicolumn{5}{|l|}{Weibull survival distribution} \\
$\alpha_{nt}$     & 0.670   &    0.710          &  $-$0.700   & 0.680\\
$\alpha_{t}$      & 1.030   &    0.980         &  0.990      & 0.970\\
$\beta_{nt,0}$    & $-$2.680&    $-$2.750     &  $-$2.740   & $-$2.720\\
$\beta_{nt,1}$    & 0.070   &      0.260       &  0.260      & 0.050\\
$\beta_{t,0}$     & $-4.380$     &  $-$4.300     &  $-$4.250   & $-$4.230\\
$\beta_{t,1}$     &1.370    &      1.330       &   1.390    & 1.350 \\
$b_0$             &0.450   &   0.550         &  3.540      & $-$1.770\\
$b_1$             &0.280   &    0.740       &   4.140    & 1.060 \\
\hline
 \multicolumn{5}{|l|}{Gompertz survival distribution} \\
$\gamma_{nt}$   &0.001    &   0.004           & 0.001    & 0.001 \\
$\gamma_{t}$    &0.060    & 0.040           & 0.040     & 0.060\\
$\lambda_{nt,0}$ &$-$3.370& $-$3.450     & $-$3.420  & $-$3.370\\
$\lambda_{nt,1}$ &0.140   &0.360       &  0.350    & 0.110 \\
$\lambda_{t,0}$  &$-$4.790& $-$4.550    &  $-$4.620 & $-$4.820\\
$\lambda_{t,1}$ &1.490    & 1.460        &  1.510    & 1.490\\
$b_0$           &0.360    &  0.580           &  3.280    & $-$2.250\\
$b_1$           &0.280    &  0.900          &  4.050    &  1.310\\
\hline
\end{tabular}
\caption{True values used in simulation study 2: misspecification of the survival distribution.}
\label{table:true_values}
\end{table}

\newpage
\section{Results of Simulation Study 2}
\begin{sidewaystable}
\centering
\footnotesize
\begin{tabular}{|l | r  r r | r  r r | r r r | r r r | r r r  | }
\hline
 \multirow{2}{*}{\shortstack{Underlying \\distribution}} &   \multicolumn{3}{c|}{$\mbox{HR}_{NT}$ }   &  \multicolumn{3}{c|}{$\mbox{HR}_{T}$ }&  \multicolumn{3}{c|}{ Reference $\rho$} &  \multicolumn{3}{c|}{ Covariate $\rho$} & \multicolumn{3}{c|}{Percentage Chosen} \\
\cline{2-16}
 &    Bias & CP & MSE & Bias & CP & MSE & Bias & CP & MSE & Bias & CP & MSE & Exp & Weibull & Gompertz \\
\hline
\textbf{Normal}  &&& &&&&&&&&&&&&\\

Exponential&	0.001&	95.4&	0.000&	$-$0.018&	94.1&	0.100&	$-$0.001&	95.4&	0.000&	0.002&	95.0&	0.000&	79.8&	10.4&	9.8\\
Weibull&	    0.007&	93.0&	0.000&	0.012&	    97.0&	0.100&	0.003   &	97.0&	0.000&	0.003&	97.0&	0.000&	0.0&	100.0&	0.0\\
Gompertz&	    0.066&	92.0&	0.300&	0.252&	    94.0&	0.600&	0.033&	    96.0&	0.200&	0.029&	90.0&	0.200&	0.0&	2.0&	98.0\\

\hline
\textbf{Clayton}             &&& &&&&&&&&&&&&\\

Exponential&	0.001&	93.8&	0.01&	0.014&	93.8&	0.097&	$-$0.001&	95.8&	0.002&	0&	94&	0&	78.8&	11&	10.2\\
Weibull&	$-$0.002&	94.6&	0.008&	0.009&	93.9&	0.087&	$-$0.002&	95.8&	0.002&	$-$0.001&	94.7&	0&	0&	100&	0\\
Gompertz&	0.001&	93.3&	0.011&	0.013&	94.5&	0.097&	$-$0.001&	94.6&	0.002&	0&	94.2&	0&	0&	3.1&	96.9\\

\hline
\textbf{Frank}         &&& &&&&&&&&&&&&\\

Exponential&	0.002&	94.3&	0.010&	0.011&	94.4&	0.094&	0.000&	94.5&	0.002&	$-$0.001&	94.1&	0.000&	82.4&	14.5&	3.1\\
Weibull&	$-$0.002&	94.1&	0.008&	0.009&	94.6&	0.09&	0.000&	94.9&	0.002&	$-$0.001&	95.1&	0.000&	0.0&	100.0&	0.0\\
Gompertz&	$-$0.014&	92.4&	0.010&	0.031&	93.9&	0.123&	0.033&	92.1&	0.003&	$-$0.015&	89.2&	0.001&	0.0&	8.7&	91.3\\
\hline

\textbf{Gumbel}              &&& &&&&&&&&&&&&\\

Exponential&	0.168&	84.8&	0.27&	0.018&	80.2&	0.091&	0.001&	95.0&	0.001&	$-$0.003&	95.0&	0.002&	79.5&	9.6&	10.9\\
Weibull&	$-$0.002&	95.1&	0.006&	0.021&	94.5&	0.085&	0.001&	96.0&	0.001&	$-$0.002&	94.9&	0.001&	0.1&	99.8&	0.1\\
Gompertz&	$-$0.003&	95.7&	0.006&	0.017&	94.9&	0.109&	0.002&	96.5&	0.001&	$-$0.004&	93.8&	0.002&	0.0&	0.2&	99.8\\
\hline

\end{tabular}
\caption{Simulation study for the misspecification of survival distributions. Results of the simulation study with data generated from either the Exponential, Weibull or Gompertz survival distributions and the Normal, Clayton, Frank and Gumbel copulas. The model with the lowest AIC is chosen and results are recorded for 1000 data sets with 3000 individuals in each. The MSE refers to the mean squared error and CP refers to the coverage probability, given as a percentage.}
\label{post_aic_donor}
\end{sidewaystable}

\clearpage
\bibliographystyle{apa}
\bibliography{PhD}